\documentclass[journal]{IEEEtran}

\usepackage{cite}
\usepackage{bm}
\usepackage{amsmath}
\usepackage{graphicx}
\usepackage{amssymb}
\usepackage{multirow}
\usepackage{array}
\usepackage{color}
\usepackage{epsfig}
\usepackage{algorithmic}
\usepackage{graphics} 
\usepackage{epsfig} 
\usepackage{float}
\usepackage{subfigure}
\usepackage[table]{xcolor}
\allowdisplaybreaks

\renewcommand\arraystretch{1.5} 
\newtheorem{definition}{Definition}
\newtheorem{theorem}{Theorem}

\newtheorem{proposition}{Proposition}
\newtheorem{lemma}{Lemma}

\newtheorem{corollary}{Corollary}
\newtheorem{example}{Example}
\newtheorem{remark}{Remark}

\newcommand{\Lfteqn}{\begin{eqnarray} \begin{array}{lllllll}}
\newcommand{\Ndeqn}{\end{array}  \end{eqnarray}}

\begin{document}
	
\title{Opacity of Discrete Event Systems under Sensor Attacks}

\author{Shengbao Zheng, Yu Liu,Shaolong Shu,and Feng Lin
		\thanks{The authors are supported by the Doctoral Research Start-up Project of Liaoning Provincial Natural Science Foundation 2026-BS-0328 and by the National Natural Science Foundation of China under Grants 62473289.}
		\thanks{Shengbao Zheng is with the School of Automation and Electrical Engineering , Shenyang Ligong University, Shenyang 110159, Liaoning, China(E-mail: Zheng\_shengbao@sylu.edu.cn).}
		\thanks{Yu Liu is with the School of Computer Science and Technology, Tongji University, Shanghai 200092, China (E-mail: liu\_y@tongji.edu.cn).}
		\thanks{Shaolong Shu is with the School of Electronics and Information Engineering, Tongji University, Shanghai 200092, China (E-mail: shushaolong@tongji.edu.cn).}
		\thanks{Feng Lin is with the Department of Electrical and Computer Engineering, Wayne State University, Detroit, MI 48202 USA (E-mail: flin@wayne.edu).}}

\maketitle
\thispagestyle{empty}
\pagestyle{empty}

\begin{abstract}
	
	Many information control problems in discrete event systems can be addressed using strong opacity, weak opacity, and non-opacity. These opacities depend on observations. If cyber attackers can attack sensors (or observation channels), then opacities will change. An opaque system may become not opaque, or a not opaque system may become opaque. 
	Therefore, it is important to investigate opacities under sensor attacks. In this paper, we consider opacities under sensor attacks for two settings. 
	The first one is the user has prior knowledge of sensor attacks in which the opacities are called as $W$-$CA$-opacities. 
	The second one is the user has no prior knowledge of sensor attacks, and opacities are called as $N$-$CA$-opacities. 
	For the former, we construct an augmented automaton that integrates the information of sensor attacks and then converts $W$-$CA$-opacities into classical opacities to solve the verification problem. 
	For the latter, we construct the observers for the given system without sensor attacks and with sensor attacks, respectively. We then do the parallel operation on these two observers to obtain a method of verifying $N$-$CA$-opacities. Furthermore, by adopting the detector technique, we find polynomial algorithms for verifying weak opacity and non opacity for both settings.
	
	{\bf Keywords:}  Discrete event systems, opacity, security, cyber attacks, sensor attacks, automata
	
\end{abstract}

\section{Introduction}

In a discrete event system (DES), some information such as the current state and the occurred events needs to be kept secret from some adversary users but be communicated to some friendly users. 
How to control information flows in DES can be investigated by opacity. Intuitively, to avoid exposing secrets to adversary users, we would like to ensure that opacity is satisfied. 
On the other hand, to ensure the necessary information can be transferred to a friendly user such as a supervisor or a diagnoser, we would like to ensure that non-opacity (transparency) is satisfied.

Because of the above, opacity has been investigated extensively in DES as a technique for describing security properties \cite{mazare2004using, bryans2006opacity}. These studies can be divided into two categories. 
One is language-based opacity where a secret is defined as a regular set of trajectories \cite{badouel2007concurrent, lin2011opacity, zhang2015maximum, ben2016opaque, behinaein2019optimal, habbachi2022secret, dubreil2010supervisory}. The problem of whether a user can determine if a trajectory belongs to some secret languages is investigated. 
The other is state-based opacity where a secret is defined as a subset of the state set \cite{saboori2007notions, yin2019opacity,yang2021opacity, yin2018verification, tong2018current, ji2018enforcement, wu2014synthesis, wu2015synthesis, ji2019opacity, mohajerani2019compositional,YangJ2021opacity}. 
Depending on the time when the user determines the states, initial-state opacity, current-state opacity, and initial-and-final-state opacity are proposed and investigated. 
In \cite{wu2013comparative}, relationships between language-based opacity and state-based opacity are investigated and show they have the same ability to describe the security properties. 

When opacity does not hold, controlling the system to achieve opaque is meaningful. \cite{dubreil2010supervisory} investigates the problem of designing a controller to enforce opacity. 
An effective algorithm for computing the most permissive controller to ensure opacity is also proposed. \cite{tong2018current} investigates opacity enforcement problem using supervisory control theory.  A locally optimal supervisor is designed to enforce current-state opacity. The opacity enforcement problem is also investigated by the ``insertion function" approach \cite{ji2018enforcement, wu2014synthesis, wu2015synthesis} and the ``edit function" approach \cite{ji2019opacity, mohajerani2019compositional}. 
In \cite{habbachi2022secret}, within a Petri net framework, the authors investigate the opacity enforcement problem by sensor-failure-making strategy (add some events) and sensor-injection-making strategy (delete some events). 
For surveys and overviews of the opacity of discrete event systems, the reader is referred to \cite{jacob2016overview, lafortune2018history, basilio2021analysis}.

\begin{figure}[htb]
		\centering
		\includegraphics[scale=0.8]{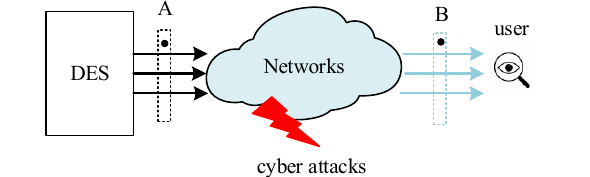}
		\caption{The observation of the user under cyber attacks} 
		\label{fig:Opacity-CA}
\end{figure}

Most of results in the literature assume that the observation of the user is the output of the given system, that is, the observation at point $A$ in Fig.  \ref{fig:Opacity-CA}. 
However, in a network environment, this assumption may not be true anymore \cite{yang2021opacity}, \cite{chen2022attackable}, \cite{lin2023diagnosability}. 
The user observes event occurrences via some communication networks and can only collect the observation at point $B$, not at point $A$. Due to cyber attacks in networks, observations at point $B$ and at point $A$ are different. 
The difference of observations caused by cyber attacks changes the opacity property. We call the attacks on observations as sensor attacks and re-consider the opacity problem under cyber attacks.

Cyber attacks have been investigated in DES with different attack models \cite{wakaiki2019supervisory, meira2020synthesis, meira2021synthesisS, meira2021synthesis, alves2022discrete, rashidinejad2019supervisory, lima2021security, su2018supervisor}, including ``the sensor deception attacks",  ``the man-in-middle attacks", and ``the bounded sensor reading alterations". In this paper, we adapt the attack model proposed in \cite{9501323,zheng2021modeling} to investigate opacity.
We name the attack model in \cite{9501323,zheng2021modeling} as the ALTER model  (for Attack Language, Transition-Based, Replacement), where an attacker can replace the event in an attackable transition by one of the strings in the corresponding attack language.
The attack languages can be obtained from the knowledge of the attacker. Because of the flexibility of the language representation, insertion attacks, deletion attacks, and replacement attacks can all be modeled using attack languages. 

Depending on whether the user knows the ALTER model or not, there are two possible settings to be investigated. In the first setting, the user has prior knowledge of sensor attacks. In this case, the user receives the observations at point B and knows the observations have been changed by sensor attackers. 
In the second setting, the user has no prior knowledge of sensor attacks. In this case, the user receives the observations at point B and mistakenly believes the observations are received at point A. 
Both settings have been investigated in \cite{chen2022attackable, zhang2021joint, li2022tamper, alves2022robust, li2021error, zhang2023robust} using other attack models. 

In this paper, we investigate the opacity problem under sensor attacks for these two settings. For a DES modeled by an automaton $G$, we define the opacities under sensor attacks for the setting in which the user has prior knowledge of sensor attacks as $W$-$CA$-opacities. 
We define the opacities under sensor attacks for the setting in which the user has no prior knowledge of sensor attacks as $N$-$CA$-opacities. 
For the former, we replace each attackable transition in $G$ with the corresponding automaton in the ALTER model and obtain a new automaton $G^a$. We prove that $G$ is $W$-$CA$-opaque if and only if $G^a$ is opaque. Thus, we convert $W$-$CA$-opacities into classical opacities and then solve the opacity verification problem using the existing methods. 
For the latter, we construct two observers, one for the system without sensor attacks and the other for the system with sensor attacks. We perform the parallel operation on these two observers to obtain an augmented automaton that includes all state estimates for the observations at point A and point B, respectively. We then propose a method to verify $N$-$CA$-opacities based on the augmented automaton. Furthermore, by adopting the detector technique, we find polynomial algorithms for verifying weak opacity and non opacity for both settings. 

The paper is organized as follows. In Section \ref{S2}, we review the language-based opacities of DES and the ALTER model. In Section \ref{s3}, we define opacities of DES under sensor attacks for two different settings as $W$-$CA$-opacities and $N$-$CA$-opacities.
In Section \ref{s4}, we convert $W$-$CA$-opacities into classical opacities to solve its verification problem. In Section \ref{s5}, we develop a method to check $N$-$CA$-opacities by combining two observers. 
In Section \ref{s6}, we develop polynomial algorithms for checking weak/non-$W$-$CA$-opacity and  weak/non-$N$-$CA$-opacity using the detector technique. Finally, we conclude the paper in Section \ref{s7}.

\section{Background}\label{S2}

\subsection{Opacity of Discrete Event Systems}

A discrete event system is modeled by a (deterministic) automaton \cite{cassandras2021introduction}:
$$
G=(Q, \Sigma, \delta, q_0, Q_m),
$$
where $\Sigma$ is the set of events; $Q$ is the set of states; $\delta : \ Q \times \Sigma \rightarrow Q$ is the transition function; $q_0 \in Q$ is the initial state; $Q_m$ is the set of marked states. The transition function can be extended to $\delta: \ Q \times \Sigma ^* \rightarrow Q$ in the usual way. For a string $s \in \Sigma ^*$, we use $\delta(q_0, s) !$ to denote that $\delta(q_0, s)$ is defined. The active event function $\Gamma: Q\rightarrow 2^\Sigma$ is defined as follows. For a state $q\in Q$, 
$$
\Gamma(q) = \{ \sigma \in \Sigma : \delta(q, \sigma)! \}.
$$
The set of all strings that are defined from the initial state is called the language generated by $G$, which is defined as:
$$
L(G) = \{ s \in \Sigma^{*} : \delta(q_0, s)!\}.
$$
The set of all strings that start at the initial state and end in a marked state is called the language marked by $G$, which is defined as:
$$
L_m(G) = \{ s \in L(G) : \delta(q_0, s) \in Q_m\}.
$$

In general, a language $K \subseteq \Sigma ^*$ is a set of strings. For a string $s \in \Sigma ^*$, we say $s'$ is a prefix of $s$ if $(\exists s'' \in \Sigma^*)s's''=s$ and we say $s'$ is a suffix of $s$ if $(\exists s'' \in \Sigma^*)s''s'=s$. The (prefix) closure of $K$, denoted by $\overline{K}$, is the set of prefixes of strings in $K$. 
A language is (prefix) closed if it equals its prefix closure. By the definition, $L(G)$ is closed. 

For a string $s \in \Sigma ^*$, we use $|s|$ to denote its length. For a set $Q' \subseteq Q$, we use $|Q'|$ to denote its cardinality.

The set of events $\Sigma$ is partitioned into observable event set $\Sigma_o$ and unobservable event set $\Sigma_{uo}$, that is, $\Sigma = \Sigma _o \dot\cup \Sigma _{uo}$. An external user can only observe the occurrences of events in $\Sigma _o$. The observation mapping can be described by the natural projection $P: \Sigma ^* \rightarrow \Sigma _o^*$, which is defined recursively as:
\begin{align*}
	P(\varepsilon) = \varepsilon; \ \ 	P(s\sigma) = \left\{ \begin{array}{ll}
		P(s) \sigma & \mbox{if } \sigma \in \Sigma _o \\
		P(s) & \mbox{if } \sigma \in \Sigma _{uo}
	\end{array} \right.,
\end{align*}
where $s \in \Sigma ^*$, $\sigma \in \Sigma$, and $\varepsilon$ denotes the empty string. The inverse of projection $P$ is denoted by $P^{-1}$ and defined as follows. For $v \in \Sigma _o^*$,
$$
P^{-1}(v) = \{ s \in L(G) : P(s)=v \}.
$$
Both $P$ and $P^{-1}$ can be extended from a string to a language in the usual way \cite{ cassandras2021introduction}.

Let $J, K \subseteq L(G)$ be two languages. We define opacities as follows \cite{lin2011opacity} \footnote{Without loss of generality, we adopt language-based definitions proposed in \cite{lin2011opacity} for opacities in this paper.}.

\begin{definition}		
	For a discrete event system $G$ and two languages $J, K \subseteq L(G)$, 
	$J$ is strongly opaque with respect to $K$
	and $P$ if
	$$
	J \subseteq P^{-1}P(K).
	$$
	$J$ is weakly opaque with respect to $K$ and $P$ if
	$$
	J \cap P^{-1}P(K) \not= \emptyset.
	$$
	$J$ is not opaque with respect to $K$ and $P$ if
	$$
	J \cap P^{-1}P(K) = \emptyset.
	$$
\end{definition}

In a discrete event system, after the occurrence of $s \in L(G)$, the observation $P(s)$ is unique. 
We then have the following results \cite{wu2013comparative}.

\begin{proposition}\label{p of euqivalent strongly opcaque}
	Consider a mapping $\eta : \Sigma ^* \rightarrow \Sigma ^*$, where $\eta (s)$ is unique. The following three statements are equivalent.	
	\begin{enumerate}
		\item
		$J \subseteq \eta^{-1}\eta(K)$.
		\item
		$(\forall s \in J) (\exists s' \in K) \eta (s) = \eta (s')$.
		\item
		$\eta(J)\subseteq \eta(K)$.
	\end{enumerate}
\end{proposition}

\begin{proposition}\label{p of euqivalent weakly opcaque}
	Consider a mapping $\eta : \Sigma ^* \rightarrow \Sigma ^*$, where $\eta (s)$ is unique. The following three statements are equivalent.
	\begin{enumerate}
		\item
		$J \cap \eta^{-1}\eta(K) \not= \emptyset$.
		\item
		$(\exists s \in J) (\exists s' \in K) \eta (s) = \eta (s')$.
		\item
		$\eta(J)\cap \eta(K) \not= \emptyset$.
	\end{enumerate}	
\end{proposition}

By Propositions \ref{p of euqivalent strongly opcaque} and \ref{p of euqivalent weakly opcaque}, we have the following corollary.
\begin{corollary}\label{CheOpa}
	Consider a DES under partial observation. $J$ is strongly opaque with respect to $K$ and $P$ if and only if
	$$
	P(J)\subseteq P(K).
	$$
	$J$ is weakly opaque with respect to $K$ and $P$ if and only if
	$$
	P(J) \cap P(K) \not= \emptyset.
	$$	
	$J$ is not opaque with respect to $K$ and $P$ if and only if
	$$
	P(J) \cap P(K) = \emptyset.
	$$
\end{corollary}

In order to verify the conditions in Corollary \ref{CheOpa}, without loss of generality, we assume that $J$ and $K$ are marked by some states $Q_J$ and $Q_K$ in $G$, that is,
\begin{align*}
	J=L_m(G,Q_J) &= \{ s \in L(G) : \delta(q_0, s) \in Q_J \} \\
	K=L_m(G,Q_K) &= \{ s \in L(G) : \delta(q_0, s) \in Q_K \}.
\end{align*}
Here we extend the definition of $L_m(G)$ to $L_m(G,Q_J)$ and $L_m(G,Q_K)$ to indicate different sets of marked states. We replace $Q_m$ in $G$ by $Q_J$ and $Q_K$ as
$$
G=(Q, \Sigma, \delta, q_0, Q_J, Q_K) .
$$ 

We construct the observer for $G$ as 
\begin{align*}
	G_{obs}&=OBS(G)=(X,\Sigma_o,\xi, x_0, X_J,X_K)\\
	&=Ac(2^Q,\Sigma_o, \xi, UR(\{q_0\}), X_J,X_K ),
\end{align*}
where $Ac(\cdot)$ denotes the accessible part. $UR(\cdot)$ is the unobservable reach defined, for $x\in 2^Q$, as 
$$
UR(x)=\{q\in Q:(\exists q'\in x)(\exists s\in \Sigma_{uo}^*)q= \delta(q',s)\}.
$$
The transition function $\xi$ is defined, for $x\in X$, and $\sigma\in \Sigma_o$, as
$$
\xi(x,\sigma)=UR(\{q\in Q:(\exists q' \in x)q= \delta(q', \sigma)\}).
$$
The marked states are defined as
\begin{equation}\label{Equation of XK}
	\begin{aligned}
		X_J&=\{x\in X:x\cap Q_J\neq \emptyset \}\\
		X_K&=\{x\in X:x\cap Q_K\neq \emptyset \}.
	\end{aligned}
\end{equation}

The following theorem, proved in \cite{lin2011opacity}, can be used to verify opacities.
\begin{theorem}\label{Throrem of opacity}
	Consider a DES $G$, observation mapping $P$ and two languages $J=L_m(G,Q_J)$,  $K=L_m(G,Q_K)$.
	\begin{enumerate}
		\item
		(Checking strong opacity) $J$ is strongly opaque with respect to $K$ and $P$ if and only if in $G_{obs}$,
		$$
		X_J \subseteq X_K.
		$$
		\item
		(Checking weak opacity) $J$ is weakly opaque with respect to $K$ and $P$ if and only if in $G_{obs}$,
		$$
		X_J \cap X_K \neq \emptyset.
		$$
		\item
		(Checking non-opacity) $J$ is not opaque with respect to $K$ and $P$ if and only if in $G_{obs}$,
		$$
		X_J \cap X_K = \emptyset.
		$$
	\end{enumerate}	
\end{theorem}

\subsection{Sensor attacks}\label{S3}

To describe sensor attacks, we use the ALTER model proposed in \cite{9501323,zheng2021modeling}, which is reviewed as follows. With a slight abuse of notation, the set of all possible transitions of $G$ is also denoted by $\delta$, that is, $\delta = \{ (q, \sigma, q'): \delta (q, \sigma) =q' \}$. 
The set of observable transitions is denoted by $\delta_o = \{ (q, \sigma, q')\in \delta : \sigma \in \Sigma _o\}$ and the set of unobservable transitions is denoted by $\delta_{uo} = \{ (q, \sigma, q') \in \delta : \sigma \in \Sigma _{uo}\}$.

We assume that some observable events/transitions can be altered by attackers in the sense that an attacker can alter these events in the observation channel. 
We denote the set of observable events and transitions that may be attacked by $\Sigma _o^a \subseteq \Sigma _o$ and $\delta ^a = \{ (q, \sigma, q') \in \delta : \sigma \in \Sigma _o^a \}$, respectively. 
We call events $\sigma \in \Sigma ^a$ attackable events and transitions ${tr}=(q, \sigma, q')\in \delta ^a$ attackable transitions.

For an attackable transition $tr=(q, \sigma, q')\in \delta ^a$, we assume that an attacker can change the observation from event $\sigma$ to any string in a language $A_{tr} \subseteq \Sigma_o ^*$. 
We denote the set of all such languages as $\mathbb{A}$ 
$$
\mathbb{A}=\{A_{tr}:tr=(q, \sigma, q')\in \delta ^a\}.
$$	
Sensor attacks are then modeled by a mapping from the set of attackable transitions to the set of languages $\mathbb{A}$ 
$$
\pi : \delta ^a\rightarrow \mathbb{A},
$$
where $\pi(tr)=A_{tr}$. Note that $\pi$ is transition-based. Note also that this general model allows for nondeterministic attacks and includes cases such as no attacks, deletion attacks, replacement attacks, and insertion attacks. For more details, the reader is referred to \cite{zheng2021modeling}.

Because attacks are nondeterministic, the observation after attacks will also become nondeterministic. For a string $s = \sigma _1 \sigma _2 ..., \sigma _{|s|} \in L(G)$ occurred in $G$, the set of possible strings after attacks, denoted by $\Theta ^a(s)$, is obtained as follows.
\begin{align*}
	\Theta ^a(s) = & \{ v_1 v_2 ..., v_{|s|}: (\forall i =
	1,2,...,|s|)(\sigma _i \in \Sigma _o^a \\
	& \wedge v_i \in A_{(\delta(q_0,\sigma_1,\cdots,\sigma_{i-1}), \sigma_i,\delta(q_0,\sigma_1,\cdots,\sigma_i))}) \\
	&\vee (\sigma _i \not\in \Sigma _o^a\wedge v_i = \sigma _i)\}.
\end{align*}

Note that $\Theta ^a(s)$ contains more than one string. Hence, $\Theta ^a$ is a mapping from $L(G)$ to $2^{\Sigma ^*}$:
$$
\Theta ^a: L(G) \rightarrow 2^{\Sigma ^*}.
$$
The observation under sensor attacks and partial observation is then given by
\begin{equation}
	\begin{split}
		\label{Phia}
		\Phi ^a (s) = P(\Theta ^a (s)).
	\end{split}
\end{equation}

Clearly, $\Phi ^a$ is a mapping from $L(G)$ to $2^{\Sigma _o^*}$. Let $(\Phi ^a)^{-1}$ be the inverse mapping of $\Phi ^a$, that is, for $v \in \Phi ^a (L(G))$,
\begin{equation}
	\begin{split}
		\label{Phi verse}
		(\Phi ^a)^{-1} (v) = \{ s \in L(G) : v \in \Phi ^a (s) \}.
	\end{split}
\end{equation}
We extended $\Phi ^a$ and $(\Phi ^a)^{-1}$ from strings to languages in the usual way as follows.
\begin{align*}
	\Phi ^a (L) & = \{ v \in \Sigma _o^* : (\exists s \in L) v \in \Phi ^a (s) \} \\
	(\Phi ^a)^{-1} (L) & = \{ s \in L(G) : (\exists v \in L) v \in \Phi ^a (s) \} .
\end{align*}


\section{Opacity of discrete event systems under sensor attacks}\label{s3}

As shown in Fig. \ref{fig:Opacity-CA}, sensor attacks change the observation of the output and hence change the opacities. Let us use an example to acquire some intuitive knowledge. 

\begin{example}\label{Example 1}
Consider the DES $G=(Q, \Sigma, \delta, q_0, Q_J, Q_K)$ shown in Fig. \ref{fig of G}. 
The state set is $Q=\{1, 2, 3, 4, 5\}$. The event set is $\Sigma = \{a, b, c, d, e,f\}$. The initial state is $q_0 = \{1\}$. The marked states are $Q_J=\{5\}$ and $Q_K=\{4\}$.
\begin{figure}[htb]
	\centering
	\includegraphics[scale=0.9]{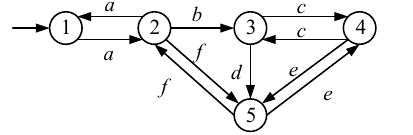}
	\caption{DES $G$}
	\label{fig of G}
\end{figure}
	
Assume that the observable events are $\Sigma _o = \{a, c, d, e, f\}$ and the events that can be attacked are $\Sigma _o^a = \{c\}$. Then we have $\delta^a=\{(3, c, 4),(4, c, 3)\}$.
	
The ALTER model is given as follows. Let $tr_1=(3, c, 4)$, $tr_2=(4, c, 3)$, $A_{tr_1}=A_{tr_2}=\{c, f\}$. In other words, the attack can replace $c$ by $f$ or do nothing. 
For $s=abc$, we have $\Theta^a(s)=\{ab\}A_{tr_1}=\{abc,abf\}$ and $\Phi^a(s)=P(\Theta^a(s))=P({\{abc,abf\}})=\{ac,af\}$. For $t\in \Phi^a(s)=af$ we can have $(\Phi^a)^{-1}(t)=\{abc,af\}$  by Equation (\ref{Phi verse}).

We consider opacities when there are no sensor attacks, that is, the observation point is A in Fig. \ref{fig:Opacity-CA}. Any string in $J$ ends with a suffix of $f$, $d$ or $ce$. Any string in $K$ ends with a suffix of $c$, $de$ or $fe$. 
Because $f$, $c$, $d$ and $e$ are all observable, strings in $J$ are distinguishable from strings in $K$ and we have $P(J)\cap P(K)=\emptyset$. Without sensor attacks, by Corollary \ref{CheOpa}, we know that $J$ is not opaque with respect to $K$ and $P$. 

Now we consider opacities when there are sensor attacks, that is, the observation point is B in Fig. \ref{fig:Opacity-CA}. Let us discuss two cases. Case 1 is when the user has prior knowledge of sensor attacks and Case 2 is when the user has no prior knowledge of sensor attacks. We take string $af$ in $J$ and string $abc$ in $K$ for consideration.

\noindent {\em Case 1:} The user has prior knowledge of sensor attacks. Hence for string $af$, we know its observation at point B is 
$$
\Phi^a(af)=\{af\}.
$$
For string $abc$, its observation at point B is 
$$
\Phi^a(abc)=\{af,ac\}.
$$
We find $\Phi^a(af)\cap \Phi^a(abc)=\{af\}\not = \emptyset$. String $af$ is indistinguishable from string $abc$.  

\noindent {\em Case 2:}  The user has no prior knowledge of sensor attacks. Since the observation is at point B, for string $af$, the user sees 
$$
\Phi^a(af)=\{af\}.
$$
However, since the user has no prior knowledge of sensor attacks, it assumes $af$ is observed at point A. He/she will compare the observation with the observation of $K$ at point A to determine whether a string in $J$ occurs or not. 
As discussed previously, at point A, we will see the occurrence of event $c$ or $e$ when a string in $K$ occurs. Hence we can distinguish string $af$ in $J$ with strings in $K$. 

\end{example}

With this knowledge, let us extend the definitions of opacities without sensor attacks (classical opacities) to the opacities under sensor attacks for the above two cases. 
In order to distinguish these two cases, we call the modified opacities for the case when the user has prior knowledge of sensor attacks as $W$-$CA$-opacities and call the modified opacities for the case when the user has no prior knowledge of sensor attacks as $N$-$CA$-opacities. 

$W$-$CA$-opacities are defined as follows.

\begin{definition}($W$-$CA$-opacity)
For a DES $G$ under sensor attacks, an observation mapping $\Phi ^a$ with respect to the ALTER model $\pi$, and two languages $J, K \subseteq L(G)$,
\begin{enumerate}
 	\item
 	$J$ is strongly $W$-$CA$-opaque with respect to $K$ and $\Phi ^a$ if all strings in $J$ are confused with some strings in $K$, that is,
	$$
	\Phi^a(J) \subseteq \Phi ^a(K).
	$$
	
	\item	
	$J$ is weakly $W$-$CA$-opaque with respect to $K$ and $\Phi ^a$ if some strings in $J$ are confused with some strings in $K$, that is,
	$$
		\Phi^a(J) \cap \Phi ^a(K) \not= \emptyset.
	$$
	
	\item
	$J$ is not $W$-$CA$-opaque with respect to $K$ and $\Phi ^a$ if any string in $J$ is not confused with any string in $K$, that is,
	$$
		\Phi^a(J) \cap \Phi ^a(K)  = \emptyset.
	$$
	\end{enumerate}
\end{definition}

$N$-$CA$-opacities are defined as follows.

\begin{definition}($N$-$CA$-opacity)
	For a DES $G$ under sensor attacks, an observation mapping $\Phi ^a$ with respect to the ALTER model $\pi$, and two languages $J, K \subseteq L(G)$, 
\begin{enumerate}
	\item
	$J$ is strongly $N$-$CA$-opaque with respect to	$K$, $P$ and $\Phi ^a$ if all strings in $J$ are confused with some strings in $K$, that is,
	$$
	\Phi^a(J) \subseteq P(K).
	$$
	\item
	$J$ is weakly $N$-$CA$-opaque with respect to $K$, $P$ and $\Phi ^a$ if some strings in $J$ are confused with some strings in $K$, that is,
	$$
	\Phi^a(J) \cap P(K) \not= \emptyset.
	$$
	\item
	$J$ is not $N$-$CA$-opaque with respect to $K$ , $P$ and $\Phi ^a$ if any string in $J$ is not confused with any string in $K$, that is,
	$$
	\Phi^a(J) \cap P(K)= \emptyset.
	$$
	\end{enumerate}

\end{definition}

\begin{remark}
	For a string $s\in L(G)$, it's observation becomes nondeterministic due to sensor attacks. It destroys some well-known properties. 
	For example, $J \subseteq (\Phi^a)^{-1} \Phi^a(K) \Leftrightarrow \Phi^a(J) \subseteq \Phi^a(K)$ does not hold any more even though $J \subseteq P^{-1}P(K) \Leftrightarrow P(J) \subseteq P(K)$. 
	Hence, we define opacities under sensor attacks on observations of languages $J$ and $K$.
\end{remark}

\section{Verification of $W$-$CA$-opacities}\label{s4}

In this section, we consider how to check $W$-$CA$-opacities. Let us assume that $A_{tr}$ is marked by an automaton $G_{tr}$. In other words, $A_{tr} = L_m (G_{tr})$ for some
$$
G_{tr} = (Q_{tr} , \Sigma, \delta_{tr}, q_{0,{tr}},
Q_{m,{tr}}).
$$

We integrate the ALTER model into the automaton $G$ by replacing each transition $tr=(q, \sigma, q') \in \delta ^a$ with $G_{tr}$ as follows.
\begin{align*}
	G_{tr\rightarrow G_{tr}} = (Q \cup Q_{tr}, \Sigma , \delta_{tr\rightarrow G_{tr}}, q_0 ),
\end{align*}
where $\delta_{tr\rightarrow G_{tr}} = (\delta - \{(q, \sigma, q')\})\cup  \delta_{tr} \cup \{ (q, \varepsilon, q_{0,tr}) \} \cup \{ (q_m, \varepsilon, q'): q_m \in Q_{m,tr} \}$. 
In other words, automaton $G_{tr\rightarrow G_{tr}}$ contains all transitions in $\delta$, except $(q,\sigma,q')$, and in $\delta_{tr}$, plus the $\varepsilon$-transitions from $q$ to the initial state of $G_{tr}$ and from marked states of $G_{tr}$ to $q'$.

Denote the automaton after replacing all attackable transitions as
$$
G^a=(Q \cup \tilde{Q}, \Sigma , \delta ^a , q_0, Q_J, Q_K ),
$$
where $\tilde{Q}= \cup_{tr \in \delta ^a} Q_{tr}$ is the set of states added during the replacement. Note that $G^a$ is nondeterministic, that is, its transition function is a mapping $\delta ^a: (Q \cup \tilde{Q}) \times \Sigma^* \rightarrow 2^ {Q \cup \tilde{Q}}$. The active event function $\Gamma: (Q \cup \tilde{Q}) \rightarrow 2^\Sigma$ is defined as follows. For a state $q\in Q \cup \tilde{Q} $, 
$$
\Gamma(q) = \{ \sigma \in \Sigma : \delta ^a (q, \sigma) \not= \emptyset \}.
$$
The language generated by $G^a$ is defined as 
$$
L(G^a) = \{ s \in \Sigma^{*} : \delta ^a (q_0, s) \not= \emptyset \}.
$$
The two marked languages with respect to $Q_J$ and $Q_K$ are defined as as 
\begin{align*}
	J^a&=L_m(G^a, Q_J)=\{s\in L(G^a):\delta ^a(q_0,s)\cap Q_J\not =\emptyset\}\\
	K^a&=L_m(G^a, Q_K)=\{s\in L(G^a):\delta ^a(q_0,s)\cap Q_K\not =\emptyset\}.
\end{align*}

We now have the following theorem to verify $W$-$CA$-opacities.
\begin{theorem} \label{Theorem of CA opacity}
	Consider a DES $G$ under sensor attacks, an observation mapping $\Phi ^a$ with respect to the ALTER model $\pi$, and two languages $J=L_m(G,Q_J)$ and $K=L_m(G,Q_K)$.
	\begin{enumerate}
		\item 
		(Check strong $W$-$CA$-opacity) $J$ is strongly $W$-$CA$-opaque with respect to $K$ and $\Phi ^a$ if and only if $J^a$ is strongly opaque with respect to $K^a$ and $P$.
		\item
		(Check weak $W$-$CA$-opacity) $J$ is weakly $W$-$CA$-opaque with respect to $K$ and $\Phi ^a$ if and only if $J^a$ is weakly opaque with respect to $K^a$ and $P$.
		\item
		(Check non-$W$-$CA$-opacity) $J$ is not $W$-$CA$-opaque with respect to $K$ and $\Phi ^a$ if and only if $J^a$ is not opaque with respect to $K^a$ and $P$.
	\end{enumerate}
\end{theorem}
\begin{IEEEproof}
	Since we replace each transition $tr\in \delta^a$ with the corresponding automaton $G_{tr}$ such that  $A_{tr} = L_m(G_{tr})$, we have, for any string $s\in \Sigma^*$,
	\begin{equation}\label{equation of Ga and Theta}
			\begin{aligned}
			&s\in L_m(G^a, Q_J)\\
			\Leftrightarrow &(\exists s'\in L_m(G, Q_J))s\in \Theta^a(s').	
		\end{aligned}
	\end{equation}
Let us first prove
\begin{equation}\label{equation of Phi(J)=P(J)}
\Phi^a(J)=P(J^a)
\end{equation}
as follows. For any $t\in \Sigma_o^*$,
\begin{align*}
	&t\in \Phi^a(J)\\
	\Leftrightarrow 
	&(\exists s\in  L_m(G, Q_J))t\in\Phi^a(s)\\
	&\text{(by the definition of $\Phi^a(J)$)} \\ 
	\Leftrightarrow 
	&(\exists s\in  L_m(G, Q_J))t\in P(\Theta^a(s))\\
	&\text{(by Equation (\ref{Phia})) }\\
	\Leftrightarrow 
	&(\exists s\in  L_m(G, Q_J))(\exists s' \in \Theta^a(s))t\in P(s')\\
	\Leftrightarrow 
	&(\exists s'\in  L_m(G^a, Q_J))t\in P(s')\\
	&\text{(by Equation (\ref{equation of Ga and Theta})) }\\
	\Leftrightarrow 
	&t\in P(J^a).
\end{align*}

Similarly, we have
\begin{equation}\label{equation of Phi(K)=P(K)}
\Phi^a(K)=P(K^a).
\end{equation}

Now let us prove Theorem \ref{Theorem of CA opacity} as follows. For the case of strong $W$-$CA$-opacity, we have
\begin{align*}
	&J \text{ is strongly $W$-$CA$-opaque with respect to $K$ and $\Phi ^a$} \\
	\Leftrightarrow 
	&\Phi^a(J)\subseteq \Phi^a(K)\\
	\Leftrightarrow 
	&P(J^a)\subseteq P(K^a)\\
	&\text{(by  Equations (\ref{equation of Phi(J)=P(J)}) and (\ref{equation of Phi(K)=P(K)}))} \\ 
	\Leftrightarrow 
	&J^a \text{ is strongly opaque with respect to $K^a$ and $P$}.
\end{align*}
 
Similarly, we can prove the cases for weak $W$-$CA$-opacity and non-$W$-$CA$-opacity.

%
\end{IEEEproof}


By Theorem \ref{Theorem of CA opacity}, $W$-$CA$-opacities can be ``converted'' to classical opacities. In particular, we can check $W$-$CA$-opacities using the methods developed for checking classical opacities as follows.

Let us construct the observer  for $G^a$ using operator $OBS$ as follows.
\begin{align*}
	G_{obs}^a & = OBS(G^a)\\
	& =(Y,\Sigma _o, \xi^a, y_0, Y_J, Y_K)\\
	&= Ac(2^{Q \cup \tilde{Q}},\Sigma _o, \xi^a, UR(\{q_0\}), Y_J, Y_K),
\end{align*}
where
$$
UR(y) = \{ q \in Q \cup \tilde{Q}: (\exists q' \in y)(\exists s\in \Sigma_{uo}^*) q\in\delta^a (q', s) \}
$$
and
$$
\xi^a (y, \sigma) = UR(\{q \in Q \cup \tilde{Q}: (\exists q' \in y) q\in\delta^a (q',\sigma)\}).
$$
The marked states are defined as
\begin{equation}\label{Equation of YJ}
\begin{aligned}
	Y_J &=\{y \in Y: y \cap Q_J \not= \emptyset \} \\
	Y_K &=\{y \in Y: y \cap Q_K \not= \emptyset \}.
\end{aligned}
\end{equation}

By Theorems \ref{Throrem of opacity} and \ref{Theorem of CA opacity}, we have the following corollary.

\begin{corollary}\label{corollary of CA}
	Consider a DES $G$ under sensor attacks, an observation mapping $\Phi ^a$ with respect to the ALTER model $\pi$, and two languages $J=L_m(G,Q_J)$ and $K=L_m(G,Q_K)$.
	\begin{enumerate}
		\item
		(Checking strong $W$-$CA$-opacity) $J$ is strongly $W$-$CA$-opaque with respect to $K$ and $\Phi^a$ if and only if in $G_{obs}^a$,
		$$
		Y_J \subseteq Y_K.
		$$
		\item
		(Checking weak $W$-$CA$-opacity) $J$ is weakly $W$-$CA$-opaque with respect to $K$ and $\Phi^a$ if and only if in $G_{obs}^a$,
		$$
		Y_J \cap Y_K \neq \emptyset.
		$$
		\item
		(Checking non-$W$-$CA$-opacity) $J$ is not $W$-$CA$-opaque with respect to $K$ and $\Phi^a$ if and only if in $G_{obs}^a$,
		$$
		Y_J \cap Y_K = \emptyset.
		$$
	\end{enumerate}		
\end{corollary}

\begin{remark}
	 It is well known that the number of states in observer $G_{obs}^a$ is upper bounded by $2^{|Q \cup \tilde{Q}|}$. Verifying $W$-$CA$-opacities using Corollary \ref{corollary of CA} need to check every state in $G_{obs}^a$ for one time. Hence the computational complexity is exponential.
\end{remark}

Let us use an example to illustrate the results.

\begin{example}\label{Example 2}
Let us consider the same DES as in Example \ref{Example 1}. We can construct $G_{tr_1}$ and $G_{tr_2}$ for $A_{tr_1}$ and $A_{tr_2}$ as shown in Fig. \ref{fig of Gtr}. Note that $A_{tr_1}=L_m(G_{tr_1})$ and $A_{tr_2}=L_m(G_{tr_2})$. The marked states of $G_{tr_1}$ and $G_{tr_2}$ are $B$ and $D$, respectively. We then obtain $G^a$ as shown in Fig. \ref{fig of Ga}.
	
\begin{figure}[htb]
	\centering
	\includegraphics[scale=0.9]{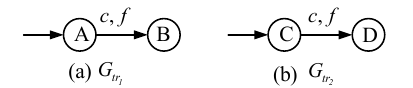}
	\caption{Sensor attack models}
	\label{fig of Gtr}
\end{figure}

\begin{figure}[htb]
	\centering
	\includegraphics[scale=0.9]{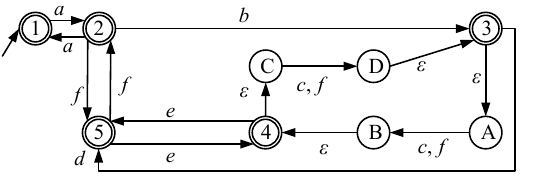}
	\caption{$G^a$}
	\label{fig of Ga}
\end{figure}

For $J=L_m(G,Q_J)$ and $K=L_m(G,Q_K)$, we can check if $J$ is strongly/weakly/not $W$-$CA$-opaque with respect to $K$ and $\Phi ^a$ by checking if $J^a$ is strongly/weakly/not opaque with respect to $K^a$ and $P$. 

In this example, $b$ is unobservable. We construct  $OBS(G^a)=G^a_{obs}$  as shown in Fig. \ref{fig of Gaobs}, where 
\begin{align*}
	Y=\{&\{1\}, \{2, 3, A\}, \{4, C\},\{4, 5, C\},\{4, 5, B, C\},\\
	&\{2, 3, A, D\},\{3, A, D\},\{4, B, C\}, \{5\}\}.
\end{align*}

\begin{figure}[htb]
	\centering
	\includegraphics[scale=0.8]{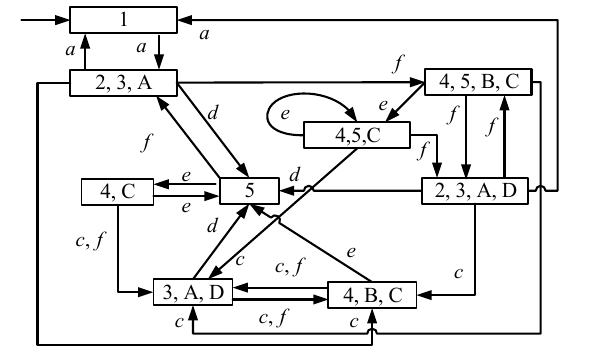}
	\caption{$G^a_{obs}$}
	\label{fig of Gaobs}
\end{figure}

Let us consider different $J$ and $K$. For $J_1=L_m(G, \{4\})$ and $K_1=L_m(G, \{5\})$, we have
\begin{align*}
	Y_{J_1}=&\{y\in Y: y\cap \{4\} \neq \emptyset \}\\
	=&\{\{4, C\},\{4, 5, C\},\{4, 5, B, C\},\{4, B, C\}\}\\
	Y_{K_1}=&\{y\in Y: y\cap \{5\} \neq \emptyset \}\\
	=&\{\{4, 5, B, C\},\{4, 5, C\},\{5\}\}.
\end{align*}
Since $Y_{J_1}\cap Y_{K_1} \neq \emptyset$, by Corollary \ref{corollary of CA}, $J_1$ is weakly $W$-$CA$-opaque with respect to $K_1$ and $\Phi ^a$.

For $J_2=L_m(G, \{2\})$ and $K_2=L_m(G, \{3\})$, we have 
\begin{align*}
Y_{J_2}=&\{y\in Y: y\cap \{2\} \neq \emptyset \}\\
=&\{\{2, 3, A\},\{2, 3, A, D\}\}\\
Y_{K_2}=&\{y\in Y: y\cap \{3\} \neq \emptyset \}\\
=&\{\{2, 3, A\},\{2, 3, A, D\},\{3, A, D\}\}.
\end{align*}
Since $Y_{J_2} \subseteq Y_{K_2}$, by Corollary \ref{corollary of CA}, $J_2$ is strongly $W$-$CA$-opaque with respect to $K_2$ and $\Phi ^a$.

For $J_3=L_m(G, \{1\})$ and $K_3=L_m(G, \{2, 3, 4, 5\})$, we have 
\begin{align*}
	Y_{J_3}=&\{y\in Y: y\cap \{1\} \neq \emptyset \}\\
	=&\{\{1\}\}\\
	Y_{K_3}=&\{y\in Y: y\cap \{2, 3, 4, 5\} \neq \emptyset \}\\
	=&\{\{2, 3, A\}, \{4, C\},\{4, 5, C\},\{4, 5, B, C\},\\
	&\{2, 3, A, D\},\{3, A, D\},\{4, B, C\}, \{5\}\}.
\end{align*}
Since $Y_{J_3}\cap Y_{K_3} = \emptyset$, by Corollary \ref{corollary of CA}, $J_3$ is not $W$-$CA$-opaque with respect to $K_3$ and $\Phi ^a$.

		
\end{example}

\section{Verification of $N$-$CA$-opacities}\label{s5}

In this section, we investigate how to verify $N$-$CA$-opacities. The definition of $N$-$CA$-opacities tells us we can verify $N$-$CA$-opacities by comparing $\Phi ^a(J)$ and $P(K)$. Let us construct automata to describe these two languages. 

As stated in Section \ref{S2}, the observer $G_{obs}$ can be used to describe $P(K)$ as follows. 
\begin{proposition}\label{proposition of P Lm}
	Consider a DES $G$ with observation mapping $P$, its observer $G_{obs}$, and a language $K=L_m(G,Q_K)$. 
	$$
	P(K)=L_m(G_{obs}, X_K).
	$$	
\end{proposition}

\begin{IEEEproof}
It is shown in \cite{cassandras2021introduction} that, for any observation $t \in P(L(G))$,
\begin{equation}\label{Equation of xi}
	\begin{aligned}
		& \xi(q_0, t)\\
		= & \{q\in Q: (\exists s\in L(G))t\in P(s) \wedge \delta(q_0,s)=q\}.
	\end{aligned}	
\end{equation}

Hence, we can prove $P(K)=L_m(G_{obs}, X_K)$ as follows.
\begin{equation*}
	\begin{aligned}
		&t\in L_m(G_{obs}, X_K)\\
		\Leftrightarrow
		&\xi(q_0, t)\in X_K\\
		\Leftrightarrow
		&\xi(q_0, t)\cap Q_K \neq \emptyset\\
		&\text{(by Equation (\ref{Equation of XK}))}\\
		\Leftrightarrow
		&(\exists s\in L(G))t\in P(s)\wedge \delta(q_0, s)\in Q_K\\
		&\text{(by Equation (\ref{Equation of xi}))}\\
		\Leftrightarrow
		&(\exists s\in K)t\in P(s)\\
		\Leftrightarrow
		&t\in P(K).
	\end{aligned}	
\end{equation*}
\end{IEEEproof}

$\Phi ^a(J)$ can be described by the observer $G_{obs}^a$ as follows.
\begin{proposition}\label{proposition of Phi Lm }
	Consider a DES $G$ under sensor attacks, an observation mapping $\Phi ^a$ with respect to the ALTER model $\pi$, and a language $J=L_m(G,Q_J)$.
	$$
	\Phi^a(J)=L_m(G_{obs}^a, Y_J)
	$$	
\end{proposition}

\begin{IEEEproof}
It is shown in \cite{9501323}, \cite{zheng2021modeling} that, for any observation $t \in \Phi^a(L(G))$,
\begin{equation}\label{Equation of xi^a}
	\begin{aligned}
		&\xi^a(y_0, t)\cap Q\\
		= & \{q\in Q: (\exists s\in L(G))t\in\Phi^a(s) \wedge \delta(q_0,s)=q\}.
	\end{aligned}	
\end{equation}

Hence, we can prove $\Phi^a(J)=L_m(G^a_{obs}, Y_J)$ as follows.
\begin{equation*}
	\begin{aligned}
		&t\in L_m(G^a_{obs}, Y_J)\\
		\Leftrightarrow
		&\xi^a(y_0, t)\in Y_J\\
		\Leftrightarrow
		&\xi^a(y_0, t)\cap Q_J \neq \emptyset\\
		&\text{(by Equation (\ref{Equation of YJ}))}\\
		\Leftrightarrow
		&(\exists s\in L(G))t\in \Phi^a(s)\wedge \delta(q_0, s)\in Q_J\\
		&\text{(by Equation (\ref{Equation of xi^a}))}\\
		\Leftrightarrow
		&(\exists s\in J)t\in \Phi^a(s)\\
		\Leftrightarrow
		&t\in \Phi^a(J).
	\end{aligned}	
\end{equation*}
\end{IEEEproof}

In order to compare strings in $\Phi ^a(J)$ with strings in $P(K)$ conveniently, we add a ``sink'' state $x_{sink}$ and additional transitions to $x_{sink}$ from states wherever events are not defined in $G_{obs}$ as follows.
\begin{align*}
	\tilde{G}_{obs}=(X\cup \{x_{sink}\}, \Sigma_o,\tilde{\xi}, X_K), 
\end{align*}
where $\tilde{\xi}=\xi \cup \{(x, \sigma, x_{sink}):x\in X \wedge \sigma\in \Sigma_o-\Gamma(x)\}\cup\{(x_{sink}, \sigma, x_{sink}) : \sigma\in \Sigma_o\}$.


Note that $L(\tilde{G}_{obs})=\Sigma_o^*$ and $L_m(\tilde{G}_{obs}, X_K)=L_m(G_{obs}, X_K)=P(K)$.

We take the parallel composition of $G^a_{obs}$ and $\tilde{G}_{obs}$ to obtain an augmented automaton as
\begin{align*}
	G_{aug}&=G^a_{obs}||\tilde{G}_{obs}=(Z,\Sigma_o, \zeta, z_0)\\
	&=Ac(Y\times(X\cup \{x_{sink}\}), \Sigma_o, \zeta, (y_0,x_0)).
\end{align*}
Denote the states of $G_{aug}$ as $z=(y,x) \in Y\times(X\cup \{x_{sink}\})$. The marked states are defined as
\begin{equation}\label{Equation of ZJK}
\begin{aligned}
	Z_J & =\{z=(y,x) \in Z: y \in Y_J\}\\
	Z_K & =\{z=(y,x) \in Z: x \in X_K\} .
\end{aligned}
\end{equation}

With $G_{aug}$, we have the following theorem for checking $N$-$CA$-opacities.
\begin{theorem}\label{Check of N-CA-opacity}
	Consider a DES $G$ under sensor attacks, an observation mapping $\Phi ^a$ with respect to the ALTER model $\pi$, and two languages $J=L_m(G,Q_J)$ and $K=L_m(G,Q_K)$,
	\begin{enumerate}
		\item
		(Check strongly $N$-$CA$-opacity)	$J$ is strongly $N$-$CA$-opaque with respect to $K$ , $P$ and $\Phi ^a$ if and only if	
		$$
		Z_J \subseteq Z_K.
		$$
		\item
		(Check weakly $N$-$CA$-opacity) $J$ is weakly $N$-$CA$-opaque with respect to $K$,  $P$ and $\Phi ^a$ if and only if	
		$$
		Z_J \cap Z_K \neq \emptyset.
		$$
		\item
		(Check non-$N$-$CA$-opacity) $J$ is not $N$-$CA$-opaque with respect to $K$,  $P$ and $\Phi ^a$ if and only if	
		$$
		Z_J \cap Z_K =\emptyset.
		$$
	\end{enumerate}
\end{theorem}

\begin{IEEEproof}
By the definition of the parallel composition \cite{cassandras2021introduction}, we have
\begin{equation} \label{LGaug}
	\begin{split}
		L(G_{aug})& = L(G^a_{obs}) \cap L(\tilde{G}_{obs}) \\
		& = L(G^a_{obs}) \cap \Sigma _o^* \\
		&= L(G^a_{obs}) .
	\end{split}
\end{equation}
Furthermore, for any $t \in L(G_{aug})$, we have
\begin{equation}\label{zeta string}
	\zeta(z_0, t)= \zeta((y_0, x_0), t)= (\xi^a(y_0,t),\tilde{\xi}(x_0,t)) .
\end{equation}

%

Let us now prove
\begin{equation} \label{LGaaug}
	\begin{split}
		& L_m(G^a_{obs}, Y_J) =L_m(G_{aug}, Z_J)\\
		& L_m(G_{obs}, X_K)\cap L(G_{aug}) =L_m(G_{aug}, Z_K).
	\end{split}
\end{equation}

$L_m(G^a_{obs}, Y_J)=L_m(G_{aug}, Z_J)$ can be proved as follows. For any $t\in \Sigma_o^*$,
\begin{align*}
	&t\in L_m(G^a_{obs}, Y_J)\\
	\Leftrightarrow
	&\xi^a(y_0,t)\in Y_J \wedge t\in L(G^a_{obs})\\
	\Leftrightarrow
	&\xi^a(y_0,t)\in Y_J \wedge t\in L(G_{aug})\\
	&(\text{by Equation (\ref{LGaug})})\\
	\Leftrightarrow
	&\xi^a(y_0,t)\in Y_J \wedge \zeta(z_0, t) \in Z \wedge t\in L(G_{aug})\\
	&(\text{by Equation (\ref{zeta string})})\\
	\Leftrightarrow
	&\zeta(z_0,t)\in Z_J \wedge t\in L(G_{aug})\\
	&(\text{by Equation (\ref{Equation of ZJK})})\\
	\Leftrightarrow
	& t\in L_m(G_{aug}, Z_J).
\end{align*}

$L_m(G_{obs}, X_K)\cap L(G_{aug}) =L_m(G_{aug}, Z_K)$ can be proved as follows. For any $t\in \Sigma_o^*$,
\begin{align*}
	&t\in L_m(G_{obs}, X_K)\cap L(G_{aug}) \\
	\Leftrightarrow 
	& \xi(x_0,t)\in X_K \wedge t\in L(G_{aug})\\
	\Leftrightarrow 
	& \tilde{\xi}(x_0, t)\in X_K \wedge  t\in L(G_{aug}) \\
	&\text{(because $\tilde{\xi}(x_0, t) \not= x_{sink} \Rightarrow \tilde{\xi}(x_0, t)=\xi(x_0,t))$}\\
	\Leftrightarrow 
	& \tilde{\xi}(x_0, t)\in X_K \wedge \zeta(z_0, t) \in Z \wedge  t\in L(G_{aug}) \\
	& \text{(\text{by Equation (\ref{zeta string})})}\\
	\Leftrightarrow
	&\zeta(z_0,t)\in Z_K \wedge t\in L(G_{aug})\\
	&(\text{by Equation (\ref{Equation of ZJK})})\\
	\Leftrightarrow	
	&t\in L_m(G_{aug}, Z_K).
\end{align*}
	

Let us now prove Theorem \ref{Check of N-CA-opacity} as follows.
\begin{enumerate}
	\item
	(Check strongly $N$-$CA$-opacity)	
	\begin{align*}
		&Z_J\subseteq Z_K\\
		\Leftrightarrow 
		&L_m(G_{aug}, Z_J) \subseteq L_m(G_{aug}, Z_K)\\
		\Leftrightarrow 
		&L_m(G^a_{obs}, Y_J) \subseteq L_m(G_{obs}, X_K)\cap L(G_{aug})\\
		&(\text{by Equation (\ref{LGaaug})})\\
		\Leftrightarrow 
		&L_m(G^a_{obs}, Y_J) \subseteq L_m(G_{obs}, X_K) \\
		&\wedge L_m(G^a_{obs}, Y_J) \subseteq L(G_{aug})\\
		\Leftrightarrow 
		&L_m(G^a_{obs}, Y_J) \subseteq L_m(G_{obs}, X_K)\\
		&\text{(because  $L_m(G^a_{obs}, Y_J)  \subseteq L(G^a_{obs}) = L(G_{aug})$}\\
		&\text{ is always true by Equation (\ref{LGaug}))}\\
		\Leftrightarrow 
		&\Phi^a(J)\subseteq P(K) \\
		&(\text{by Propositions \ref{proposition of P Lm} and \ref{proposition of Phi Lm }}) .
	\end{align*}

	\item
	(Check weakly $N$-$CA$-opacity) 
	\begin{align*}	
		&Z_J\cap  Z_K \neq \emptyset\\
		\Leftrightarrow 
		&L_m(G_{aug}, Z_J)\cap L_m(G_{aug}, Z_K)\neq \emptyset\\
		\Leftrightarrow 
		&L_m(G^a_{obs}, Y_J)\cap L_m(G_{obs}, X_K)\cap  L(G_{aug}) \neq \emptyset\\
		&(\text{by Equation (\ref{LGaaug})})\\
		\Leftrightarrow 
		&L_m(G^a_{obs}, Y_J)\cap L_m(G_{obs}, X_K)\neq \emptyset\\
		&(\text{because  $L_m(G^a_{obs}, Y_J) \subseteq L(G_{aug})$})\\
		\Leftrightarrow	
		&\Phi^a(J)\cap P(K)\neq \emptyset \\
		&(\text{by Propositions \ref{proposition of P Lm} and \ref{proposition of Phi Lm }}) .
	\end{align*}
	\item
	(Check non-$N$-$CA$-opacity) 
	
	The proof is similar to the above.
\end{enumerate}

\end{IEEEproof}

\begin{remark}
The number of states in $G_{aug}$ is upper bounded by $2^{|Q\cup\tilde{Q}|}\cdot(2^{|Q|}+1)$. Verifying $N$-$CA$-opacities needs to check every state for one time. Hence the computational complexity is of the order $O(2^{|Q\cup\tilde{Q}|+|Q|})$.
\end{remark}

\begin{example}\label{Example 3}
Let us continue Examples \ref{Example 1} and \ref{Example 2}. We construct $OBS(G)=G_{obs}$ as shown in Fig. \ref{fig of Gobs}. We obtain $\tilde G_{obs}$ as shown in Fig. \ref{fig of tildeGobs}. 
$G^a_{obs}$ is given in Example \ref{Example 2} as shown in Fig. \ref{fig of Gaobs}. We execute the parallel operation on $\tilde G_{obs}$ and $G^a_{obs}$ to get $G_{aug}=\tilde G_{obs}||G^a_{obs}$ as shown in Fig. \ref{fig of Gaug}. 
All states $z=(y,x)$ in $G_{aug}$ and $y, x$ are shown in Table \ref{Table of Z}. We can obtain $Z_{J_1}, Z_{K_1}$ (the cells are colored with red), $Z_{J_2}, Z_{K_2}$ (the cells are colored with blue) and $Z_{J_3}, Z_{K_3}$ (the cells are colored with green). 
	\begin{figure}
	\centering
	\includegraphics[scale=0.9]{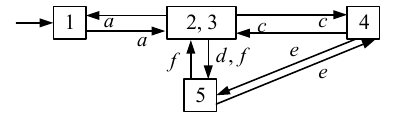}
	\caption{$G_{obs}$}
	\label{fig of Gobs}	
		\end{figure}
	\begin{figure}
		\centering
		\includegraphics[scale=0.9]{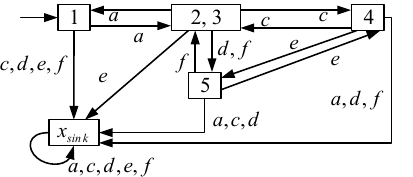}
		\caption{$\tilde G_{obs}$}
		\label{fig of tildeGobs}
	\end{figure}
	\begin{figure*}
	\centering
	\includegraphics[width=0.8\textwidth]{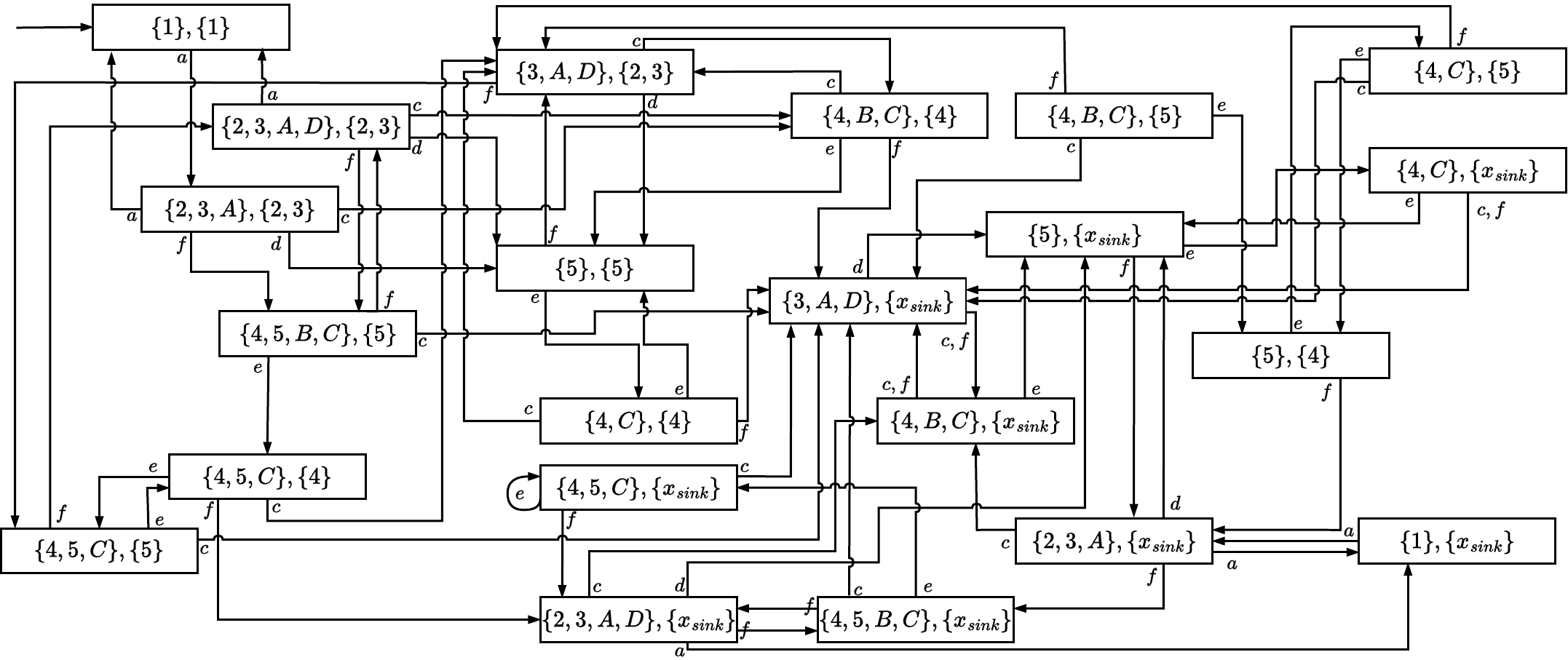}
	\caption{$G_{aug}$}
	\label{fig of Gaug}
	\end{figure*}
With $G_{aug}$ in Fig. \ref{fig of Gaug} and Theorem \ref{Check of N-CA-opacity}, we can check $N$-$CA$-opacity as follows.

%

$J_1$ is weakly $N$-$CA$-opaque with respect to $K_1$, $P$ and $\Phi ^a$ because
$$
Z_{J_1} \cap Z_{K_1} \neq \emptyset.
$$

$J_2$ is weakly $N$-$CA$-opaque with respect to $K_2$, $P$ and $\Phi ^a$ because
 $$Z_{J_2} \cap Z_{K_2} \neq \emptyset.$$
 
$J_3$ is not $N$-$CA$-opaque with respect to $K_3$, $P$ and $\Phi ^a$ because
  $$Z_{J_3} \cap Z_{K_3} = \emptyset.$$
\end{example}

\begin{table}[htb]\huge
	\centering
	\caption{All states in $G_{aug}$}
	\label{Table of Z}
	\renewcommand{\arraystretch}{1.5}
	\resizebox{0.99\columnwidth}{!}{%
	\begin{tabular}{ccccccccc}
		\hline	
		$z = (y,x) \in Z$ & $y$ & $x$ & $\in Z_{J_1}?$ & $\in Z_{K_1}?$ & $\in Z_{J_2}?$ & $\in Z_{K_2}?$ & $\in Z_{J_3}?$ & $\in Z_{K_3}?$\\
		\hline	
		$(\{1\},\{1\})$ & $\{1\}$ & $\{1\}$ & NO & NO & NO & NO & \cellcolor{green}Yes & NO\\	
		\hline
		$(\{2,3,A\},\{2,3\})$ & $\{2,3,A\}$ & $\{2,3\}$ & NO &NO &\cellcolor{blue} Yes & \cellcolor{blue}Yes & NO &\cellcolor{green} Yes\\	
		\hline
		$(\{4,B,C\},\{4\})$ & $\{4,B,C\}$ & $\{4\}$ & \cellcolor{red}Yes & NO & NO & NO & NO &\cellcolor{green} Yes\\	
		\hline
		$(\{5\},\{5\})$ & $\{5\}$ & $\{5\}$ & NO & \cellcolor{red}Yes & NO & NO & NO &\cellcolor{green} Yes\\	
		\hline
		$(\{4,5,B,C\},\{5\})$ & $\{4,5,B,C\}$ & $\{5\}$ & \cellcolor{red}Yes & \cellcolor{red}Yes & NO & NO & NO &\cellcolor{green} Yes\\	
		\hline
		$(\{3,A,D\},\{x_{sink}\})$ & $\{3,A,D\}$ & $\{x_{sink}\}$ & NO & NO & NO & NO & NO & NO\\	
		\hline
		$(\{3,A,D\},\{2,3\})$ & $\{3,A,D\}$ & $\{2,3\}$ & NO & NO & NO & \cellcolor{blue}Yes  &NO &\cellcolor{green} Yes\\	
		\hline
		$(\{4,C\},\{4\})$ & $\{4,C\}$ & $\{4\}$ &\cellcolor{red} Yes & NO & NO & NO & NO &\cellcolor{green}Yes\\	
		\hline
		$(\{5\},\{x_{sink}\})$ & $\{5\}$ & $\{x_{sink}\}$ & NO & NO & NO & NO & NO & NO\\	
		\hline
		$(\{4,B,C\},\{x_{sink}\})$ & $\{4,B,C\}$ & $\{x_{sink}\}$ &\cellcolor{red} Yes & NO & NO & NO & NO & NO\\	
		\hline
		$(\{4,B,C\},\{5\})$ & $\{4,B,C\}$ & $\{5\}$ &\cellcolor{red} Yes &\cellcolor{red} Yes & NO & NO & NO & \cellcolor{green}Yes\\	
		\hline
		$(\{4,C\},\{x_{sink}\})$ & $\{4,C\}$ & $\{x_{sink}\}$ & \cellcolor{red}Yes & NO & NO & NO & NO & NO\\	
		\hline
		$(\{2,3,A\},\{x_{sink}\})$ & $\{2,3,A\}$ & $\{x_{sink}\}$ &NO & NO &\cellcolor{blue} Yes & NO & NO & NO\\	
		\hline
		$(\{5\},\{4\})$ & $\{5\}$ & $\{4\}$ & NO & NO & NO & NO & NO &\cellcolor{green} Yes\\	
		\hline
		$(\{4,C\},\{5\})$ & $\{4,C\}$ & $\{5\}$ & \cellcolor{red}Yes &\cellcolor{red} Yes & NO & NO & NO &\cellcolor{green} Yes\\	
		\hline
		$(\{2,3,A,D\},\{2,3\})$ & $\{2,3,A,D\}$ & $\{2,3\}$ & NO &NO & \cellcolor{blue}Yes &\cellcolor{blue} Yes & NO &\cellcolor{green} Yes\\	
		\hline
		$(\{1\},\{x_{sink}\})$ & $\{1\}$ & $\{x_{sink}\}$ & NO & NO & NO & NO & \cellcolor{green}Yes & NO\\	
		\hline
		$(\{4,5,C\},\{x_{sink}\})$ & $\{4,5,C\}$ & $\{x_{sink}\}$ & \cellcolor{red}Yes & NO & NO & NO & NO & NO\\	
		\hline
		$(\{4,5,C\},\{5\})$ & $\{4,5,C\}$ & $\{5\}$ & \cellcolor{red}Yes &\cellcolor{red} Yes & NO & NO & NO &\cellcolor{green} Yes\\	
		\hline
		$(\{2,3,A,D\},\{x_{sink}\})$ & $\{2,3,A,D\}$ & $\{x_{sink}\}$ & \cellcolor{red}Yes & NO & \cellcolor{blue}Yes & NO & NO & NO\\	
		\hline
		$(\{4,5,C\},\{4\})$ & $\{4,5,C\}$ & $\{4\}$ & \cellcolor{red}Yes & NO & NO & NO & NO &\cellcolor{green} Yes\\	
		\hline
		$(\{4,5,B,C\},\{x_{sink}\})$ & $\{4,5,B,C\}$ & $\{x_{sink}\}$ &\cellcolor{red} Yes & NO & NO & NO & NO & NO\\
		\hline	
	\end{tabular}
	}
\end{table}

\section{Polynomial algorithms for verification of opacities under sensor attacks}\label{s6}

The methods in the previous section for checking $W$-$CA$-opacities and $N$-$CA$-opacities are of exponential complexity. 
In this section, we develop polynomial algorithms for checking weak  $W$-$CA$-opacity, non-$W$-$CA$-opacity, weak $N$-$CA$-opacity and non-$N$-$CA$-opacity. 
We do not have polynomial algorithms for checking strong $W$-$CA$-opacity and strong $N$-$CA$-opacity. Whether there exist such  algorithms is an open  problem.

\subsection{$W$-$CA$-opacities}

To obtain the polynomial algorithms for checking weak $W$-$CA$-opacity and non-$W$-$CA$-opacity, we construct a detector \cite{shu2011generalized} from $G^a$ using operator $DET$ as follows.
\begin{align*}
	G_{det}^a & = DET(G^a)\\
	&=(D,\Sigma _o, \rho^a, d_0, D_J,  D_K) \\
	&= Ac(D',\Sigma _o, \rho^a, UR(\{q_0\}), D_J,  D_K).
\end{align*}

In $G_{det}^a$, the initial state is $d_0 = UR(\{q_0\})$. The state set $D$ is the accessible part of $D'$, which is the initial state $d_0$ union the set of singletons and doubletons:
$$
D'=\{ d \subseteq Q \cup \tilde{Q} : |d| \leq 2 \} \cup \{ d_0 \}.
$$
The transition function $\rho^a (d, \sigma)$ is defined, for $d \in D$ and $\sigma \in \Sigma_o$, as follows. First, we define an intermediate or temporary variable
$$
Temp^a = UR(\{q \in Q \cup \tilde{Q}: (\exists q' \in d) (q', \sigma, q) \in \delta^a \}).
$$
Then according to the number of elements in $Temp$, we define
\begin{align*}
	\rho^a (d, \sigma) = \left\{ \begin{array}{ll} \{ Temp^a \} & 		\mbox{if } |Temp^a|=1 \\
		\{ d' \subseteq Temp^a : |d'|= 2 \} & \mbox{if } |Temp^a| \geq 		2 \\
		{\rm undefined } & \mbox{if } |Temp^a|=0 .
	\end{array}\right.
\end{align*}
The marked states are defined as
\begin{equation}\label{Equation of DJK}
	\begin{aligned}
		D_J & = \{d \in D: d \cap Q_J \not= \emptyset \} \\
		D_K & = \{d \in D: d \cap Q_K \not= \emptyset \}.
	\end{aligned}
\end{equation}

Note that $L(G_{det}^a)=L(G_{obs}^a)$. Detector $G_{det}^a$ has the following property \cite{shu2011generalized} .
\begin{lemma} \label{Lemma 1}
	The relation between the observer $G _{obs}^a$ and the detector $G_{det}^a$ is as follows.
	\begin{enumerate}
		\item
		For any $t \in L(G_{obs}^a)$, if $|\xi^a (y_0, t)|=1$, then $\rho^a (d_0,t) = \{ \xi^a (y_0, t) \}$.
		\item
		For any $t \in L(G_{obs}^a)$ and any $d \subseteq Q$ such that $|d|=2$, $d \in \rho^a (d_0,t) \Leftrightarrow d \subseteq \xi^a (y_0, t)$.
	\end{enumerate}	
\end{lemma}

With this property, we have the following theorem for verifying weak $W$-$CA$-opacity and non-$W$-$CA$-opacity.

\begin{theorem} \label{Theorem of Poly check Wopacity}
	Consider a DES $G$ under sensor attacks, an observation mapping $\Phi^a$ with respect to the ALTER model $\pi$, and two languages $J=L_m(G,Q_J)$ and $K=L_m(G,Q_K)$,
	\begin{enumerate}	
		\item
		(Check weak $W$-$CA$-opacity) $J$ is weakly $W$-$CA$-opaque with respect to $K$ and $\Phi ^a$ if and only if in the detector $G _{det}^a$,
		$$
		D_J \cap D_K \not= \emptyset .
		$$
		
		\item
		(Check non-$W$-$CA$-opacity) $J$ is not $W$-$CA$-opaque with respect to $K$ and $\Phi ^a$ if and only if in the $G _{det}^a$,
		$$
		D_J \cap D_K = \emptyset .
		$$
	\end{enumerate}
\end{theorem}

\begin{IEEEproof}
	Let us prove the result for weak $W$-$CA$-opacity. The proof for non-$W$-$CA$-opacity is similar. We have
	\begin{align*}
		&\text{$J$ is weakly $W$-$CA$-opaque with respect to $K$}\\
		\Leftrightarrow & Y_J \cap Y_K \neq\emptyset\\
		&\text{(by Corollary \ref{corollary of CA})}\\
		\Leftrightarrow & (\exists y \in Y) y \in Y_J \wedge y\in Y_K\\
		\Leftrightarrow & (\exists y \in Y) y \cap Q_J \neq \emptyset \wedge  y \cap Q_K \neq \emptyset\\
		&(\text{by Equation (\ref{Equation of YJ})})\\	
		\Leftrightarrow & (\exists t\in L(G_{obs}^a))\\
		&y=\xi^a(y_0,t) \wedge y \cap Q_J \neq \emptyset \wedge  y \cap Q_K \neq \emptyset\\
		&(\text{because $G^a_{obs}$ is accessible})\\
		\Leftrightarrow & (\exists t \in L(G_{obs}^a)) \\
		&\xi^a(y_0,t) \cap Q_J \neq \emptyset \wedge   \xi^a(y_0,t)\cap Q_K \neq \emptyset.
	\end{align*}
	
	We consider two possible cases as follows.
	
	\noindent {\em Case 1:} $|\xi^a (y_0, t)|=1$. In this case, by Lemma \ref{Lemma 1}, $\rho^a (d_0,t) = \{ \xi^a (y_0, t) \}$. Hence, by letting $d=\xi^a (y_0, t)$,
	\begin{align*}
		& (\exists t \in L(G_{obs}^a))  \\
		&\xi^a (y_0, t) \cap Q_J \not= \emptyset \wedge \xi^a (y_0, t) \cap Q_K \not= \emptyset \\
		\Leftrightarrow 
		& (\exists t \in L(G_{obs}^a)) \\
		& d \in \rho^a (d_0, t)\wedge d \cap Q_J \not= \emptyset \wedge d \cap Q_K \not= \emptyset\\
		\Leftrightarrow 
		& (\exists d \in D) d \cap Q_J \not= \emptyset \wedge d \cap Q_K \not= \emptyset\\
		&(\text{because $G^a_{det}$ is accessible})\\
		\Leftrightarrow 
		& (\exists d \in D) d \in  D_J  \wedge d \in D_K \\
		&(\text{because Equation (\ref{Equation of DJK})})\\
		\Leftrightarrow 
		&  D_J \cap D_K \neq \emptyset.	
	\end{align*}
	
	\noindent {\em Case 2:} $|\xi^a (y_0, t)|>1$. In this case, by Lemma \ref{Lemma 1}, $d \in \rho^a
	(d_0,t) \Leftrightarrow d \subseteq \xi^a (y_0, t)$. Hence,
	\begin{align*}
		& (\exists t \in L(G_{obs}^a))  \\
		&\xi^a (y_0, t) \cap Q_J \not= \emptyset \wedge \xi^a (y_0, t) \cap Q_K \not= \emptyset \\
		\Leftrightarrow 
		& (\exists t \in L(G_{obs}^a))(\exists q_1,q_2\in \xi^a (y_0, t))q_1\in Q_J \wedge q_2\in Q_K\\
		\Leftrightarrow 
		& (\exists t \in L(G_{obs}^a))\\
		&(\exists d\in \rho^a
		(d_0,t)) d\cap Q_J \neq \emptyset \wedge d\cap Q_K \neq \emptyset\\
		&(\text{let $q_1\in d, q_2 \in d$})\\
		\Leftrightarrow 
		& (\exists d \in D) d\cap Q_J \neq \emptyset \wedge d\cap Q_K \neq \emptyset \\
		&(\text{because $G^a_{det}$ is accessible})\\
		\Leftrightarrow 
		& (\exists d \in D) d \in  D_J  \wedge d \in D_K \\
		&(\text{by Equation (\ref{Equation of DJK})})\\
		\Leftrightarrow 
		&  D_J \cap D_K \neq \emptyset.	
	\end{align*}
\end{IEEEproof}

\begin{remark}
	It is not difficult to see that the computational complexity of checking weak $W$-$CA$-opacity and non-$W$-$CA$-opacity using Theorem \ref{Theorem of Poly check Wopacity} is of the order $O({|Q\cup\tilde{Q}|}^2)$, because the number of states in the detector is upper bounded by ${|Q\cup\tilde{Q}|}^2+ |Q\cup\tilde{Q}| +1$.
\end{remark}	

\begin{example}\label{Example 4}
	Let us continue the Example 2. We construct $DET(G^a)=G^a_{det}$ as shown in Fig. \ref{fig of G^a_{det}}. 
	Using Theorem \ref{Theorem of Poly check Wopacity}, we can check weak/non-$W$-$CA$-opacity as follows.
\begin{figure*}
	\centering
	\includegraphics[width=0.7\textwidth]{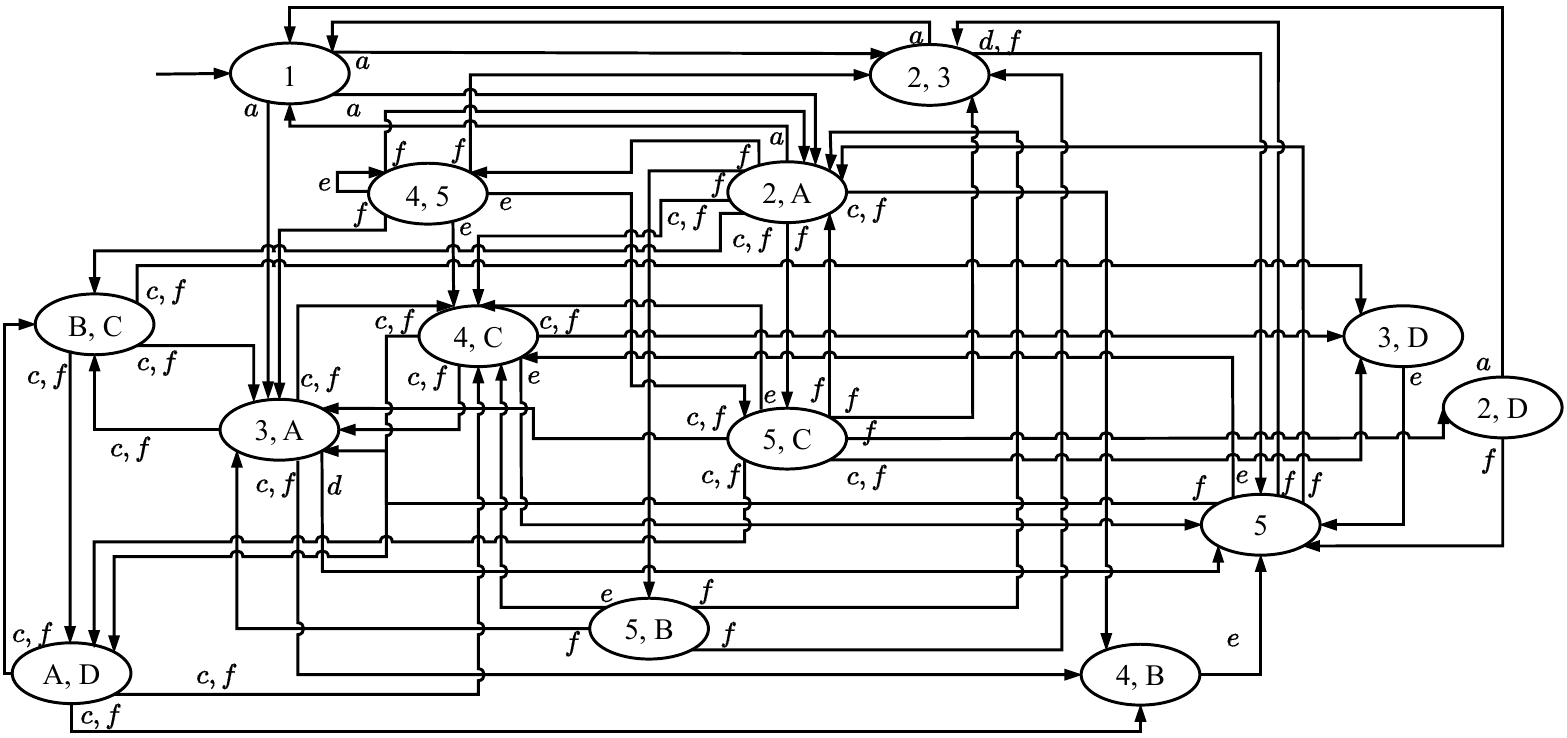}
	\caption{$G^a_{det}$}
	\label{fig of G^a_{det}}
\end{figure*}
	
	For $J_1,  J_3$ and $K_1,  K_3$  given in Example \ref{Example 2}, we have 
	\begin{align*}
		D_{J_1}=\{&\{4,5\},\{4,B\},\{4,C\}\}\\
		D_{J_3}=\{&\{1\}\}\\
		D_{K_1}=\{&\{4,5\}, \{5\}, \{5, C\}, \{5,B\}\}\\
		D_{K_3}=\{&\{2,3\},\{2,A\},\{2,D\},\{3,A\},\{3,D\},\\
		&\{4,5\},\{4,B\},\{4,C\},\{5\},\{5,C\},\{5,B\},\}.
	\end{align*} 
	
	It can be checked that 
	$
	D_{J_1}\cap D_{K_1}\neq \emptyset
	$
	and
	$
	D_{J_3}\cap D_{K_3}= \emptyset.
	$
	Hence  $J_1$ is weakly $W$-$CA$-opaque with respect to $K_1$ and $\Phi ^a$ and $J_3$ is not $W$-$CA$-opaque with respect to $K_3$ and $\Phi ^a$.
\end{example}

\subsection{$N$-$CA$-opacities}
To obtain the polynomial algorithms for checking weak $N$-$CA$-opacity and non-$N$-$CA$-opacity, we first construct the detector $G_{det}$  from $G$, using operator $DET$ as follows.
\begin{align*}
	G_{det} & = DET(G)\\
	&=(B,\Sigma _o, \rho, b_0, B_K) \\
	&= Ac(B',\Sigma _o, \rho, UR(\{q_0\}),  B_K),
\end{align*}
where $B'=\{ b \subseteq Q  : |b| \leq 2 \} \cup \{ b_0 \}.$

We define 
$$Temp = UR(\{q \in Q : (\exists q' \in b) (q',\sigma,q) \in \delta\}).$$

We then have for $b\in B$ and $\sigma\in \Sigma_o$,
\begin{align*}
	\rho(b, \sigma) = \left\{ \begin{array}{ll} \{ Temp \} &	\mbox{if } |Temp|=1 \\
		\{ b' \subseteq Temp : |b'|= 2 \} & \mbox{if } |Temp| \geq 		2 \\
		{\rm undefined } & \mbox{if } |Temp|=0.
	\end{array}\right.
\end{align*}
Define marked states as 
\begin{equation}\label{Equaiton of BK}
	\begin{aligned}
		B_K  = \{b \in B: b \cap Q_K \not= \emptyset \} .
	\end{aligned}
\end{equation}

The detector $G _{det}$ has the following property \cite{shu2011generalized} .
\begin{lemma} \label{Lemma 2}
	The relation between the observer $G_{obs}$ and  $G_{det}$ is as follows.
	\begin{enumerate}
		\item
		For any $v \in P(L(G))$, if $|\xi (x_0, t)|=1$, then $\rho (b_0,t) = \{ \xi (x_0, t) \}$.
		\item
		For any $t \in P(L(G))$ and any $d \subseteq Q$ such that $|b|=2$, $b \in \rho (b_0,t) \Leftrightarrow b \subseteq \xi (x_0, t)$.
	\end{enumerate}	
\end{lemma}

In order to include all strings in $\Sigma_o^*$, we add a state $b_{sink}$ and additional transitions to $b_{sink}$ from states wherever events are not defined in $G_{det}$ as follows.
\begin{align*}
	\tilde{G}_{det}=(B\cup\{b_{sink}\}, \Sigma_o, \tilde{\rho}, b_0, B_K),
\end{align*}
where $\tilde{\rho}=\rho\cup \{(b, \sigma, b_{sink}): b\in B \wedge \sigma \in \Sigma_o -\Gamma(b)\} \cup \{(b_{sink}, \sigma, b_{sink}): \sigma \in \Sigma_o\}.$ 
Note that $\tilde{\rho}(b, \sigma)-\{b_{sink}\}=\rho(b, \sigma)$ and $L(\tilde{G}_{det})=\Sigma_o^*$.

We take the parallel composition of $G^a_{det}$ and $\tilde{G}_{det}$ to get an augmented automaton, called $G^T_{aug}$ as
\begin{align*}
	G^T_{aug}&=G^a_{det}||	\tilde{G}_{det}=(C,\Sigma_o, \tau, c_0)\\
	&=Ac(D\times(B\cup \{b_{sink}\}), \Sigma_o, \tau, (d_0,b_0)).
\end{align*}
Denote the states of $G^T_{aug}$ as $c=(d,b) \in D\times(B\cup \{b_{sink}\})$. The marked states are defined as
\begin{equation}\label{Equation of CJK}
	\begin{aligned}
		C_J&=\{c=(d,b): d\in D_J\}\\
		C_K&=\{c=(d,b): b\in B_K\}.
	\end{aligned}
\end{equation}

For $G^T_{aug}$, we have 
\begin{align*}
	L(G_{aug}^T)& = L(G^a_{det}) \cap L(\tilde{G}_{det}) \\
	& = L(G^a_{det}) \cap \Sigma _o^* \\
	&= L(G^a_{det}) .
\end{align*}
For any $t\in L(G_{obs}^a)(=L(G_{aug}^T))$, we have
\begin{equation}\label{tau string}
	\tau(c_0, t)=\rho^a(d_0,t)\times\tilde{\rho}(b_0,t),
\end{equation}
where $c_0=(d_0, b_0)$.

We then have the following theorem.
\begin{theorem} \label{Theorem of Poly check Nopacity}
	Consider a DES $G$ under sensor attacks, an observation mapping $\Phi^a$ with respect to the ALTER model $\pi$, and two languages $J=L_m(G,Q_J)$ and $K=L_m(G,Q_K)$,
	\begin{enumerate}	
		\item
		(Check weak $N$-$CA$-opacity) $J$ is weakly $N$-$CA$-opaque with respect to $K$, $P$ and $\Phi ^a$ if and only if in 	$G^T_{aug}$,
		$$
		C_J \cap C_K \not= \emptyset.
		$$
		\item
		(Check non-$N$-$CA$-opacity) $J$ is not $N$-$CA$-opaque with respect to $K$, $P$ and $\Phi ^a$ if and only if in  $G^T_{aug}$,
		$$
		C_J \cap C_K = \emptyset.
		$$
	\end{enumerate}
\end{theorem}
\begin{IEEEproof}
	Let us prove the result for weak $N$-$CA$-opacity. The proof for non-$N$-$CA$-opacity is similar. We have 
	\begin{align*}
		&\text{$J$ is weakly $N$-$CA$-opaque with respect to $K$}\\
		\Leftrightarrow 
		&Z_J \cap Z_K \neq\emptyset\\
		&\text{(by Theorem \ref{Check of N-CA-opacity})}\\
		\Leftrightarrow 
		& (\exists z \in Z) z \in Z_J \wedge z\in Z_K\\
		\Leftrightarrow 
		& (\exists t \in L(G_{aug})) \zeta(z_0, t) \in Z_J \wedge \zeta(z_0, t) \in Z_K \\	
		&(\text{because $G_{aug}$ is accessible})\\
		\Leftrightarrow 
		& (\exists t \in L(G_{obs}^a)) \xi^a(y_0,t)\in Y_J\wedge \tilde{\xi}(x_0,t)\in X_K\\	
		&(\text{by Equations (\ref{Equation of ZJK}), (\ref{LGaug}), and (\ref{zeta string})})\\
		\Leftrightarrow 
		& (\exists t \in L(G_{obs}^a)) \xi^a(y_0,t)\cap Q_J \neq \emptyset \wedge \tilde{\xi}(x_0,t)\cap Q_K \neq \emptyset\\
		&(\text{by Equations  (\ref{Equation of XK}) and (\ref{Equation of YJ})})\\
		\Leftrightarrow 
		& (\exists t \in L(G_{obs}^a))(\exists q_1 \in\xi^a(y_0, t))(\exists q_2 \in \xi(x_0,t))\\
		&q_1\in Q_J \wedge q_2\in Q_K  \\
		&\text{(because $\tilde{\xi}(x_0, t) \not= x_{sink} \Rightarrow \tilde{\xi}(x_0, t)=\xi(x_0,t))$}\\
		\Leftrightarrow 
		& (\exists t \in L(G^T_{aug}))(\exists (d,b) \in\tau(c_0, t))\\
		&q_1 \in d \wedge q_2 \in b \wedge q_1\in Q_J \wedge q_2 \in Q_K\\
		&\text{(by Lemma \ref{Lemma 2},  $\tilde{\rho}(b, \sigma)-\{b_{sink}\}=\rho(b, \sigma)$} \\
		&\text{and Equations (\ref{tau string}))}\\
		\Leftrightarrow 
		& (\exists (d,b)\in C)d\cap Q_J \neq \emptyset \wedge b\cap Q_K \neq \emptyset\\
		&(\text{because $G^T_{aug}$ is accessible})\\
		\Leftrightarrow 
		& (\exists (d,b)\in C) d \in D_J  \wedge b \in B_K \\
		&\text{(because Equations  (\ref{Equation of DJK}) and (\ref{Equaiton of BK}))}\\
		\Leftrightarrow 
		& (\exists c\in C) c\in C_J \wedge c\in C_K \\
		&\text{(because Equation (\ref{Equation of CJK}))}\\
		\Leftrightarrow 
		&  C_J \cap C_K \neq \emptyset.
	\end{align*}
\end{IEEEproof}

\begin{remark}
	The computational complexity of checking weak $N$-$CA$-opacity and non-$N$-$CA$-opacity using Theorem \ref{Theorem of Poly check Nopacity} is of the order $O(|Q\cup\tilde{Q}|^2 \cdot|Q|^2)$ because the number of states in the augmented automaton $G^T_{aug}$ is upper bounded by $|Q\cup\tilde{Q}|^2 \cdot(|Q|^2+1)$.
\end{remark}

\begin{example}\label{Example 5}
	Let us continue the Example 3. Note that, in Fig. \ref{fig of Gobs} , for any state $x\in X$ in the observer $G_{obs} =(X,\Sigma_o, \xi, x_o)$, its cardinality is no more than 2. 
	Hence $G_{obs}=G_{det}$ and $\tilde G_{obs}=\tilde G_{det}$ (in terms of isomorphism). We take the parallel composition of $G^a_{det}$ and $\tilde G_{det}$ to get $G^T_{aug}=G^a_{det}||\tilde G_{det}$ which is partially shown in Fig \ref{fig of G^T_{aug}}.
	
	\begin{figure}
		\centering
		\includegraphics[width=\columnwidth]{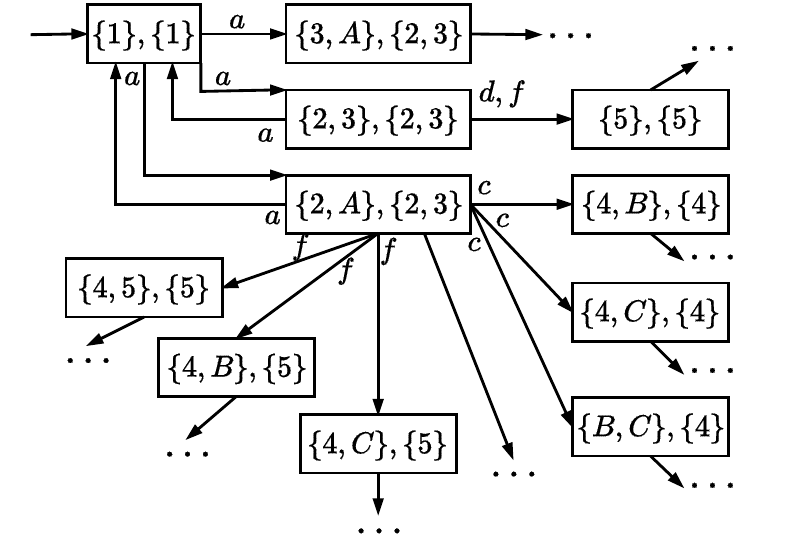}
		\caption{The part of $G^T_{aug}$}
		\label{fig of G^T_{aug}}
	\end{figure}	
	
	Let us consider $J_1$ and $K_1$  given in Example 3, by Theorem \ref{Check of N-CA-opacity}, we have checked that $J_1$ is weakly $N$-$CA$-opaque with respect to $K_1$, $P$ and $\Phi ^a$. In this example, by Equation (\ref{Equation of CJK}), we can get 
	\begin{align*}
		C_{J_1}=&\{..., \{\{4, B\}, \{4\}\}, \{\{4,C\},\{4\}\}...,\\
		&\{\{4,5\},\{5\}\},\{\{4,B\},\{5\}\},\{\{4,C\},\{5\}\},...\}\\
		C_{K_1}=&\{..., \{\{5\},\{5\}\},...,\{\{4,5\},\{5\}\},\{\{4,B\},\{5\}\},\\
		&\{\{4,C\},\{5\}\}.
	\end{align*} 
	
	It can be checked that 
	$
	C_{J_1}\cap C_{K_1}\neq \emptyset.
	$
	That is,  $J_1$ is weakly $N$-$CA$-opaque with respect to $K_1$, $P$ and $\Phi ^a$. 	
\end{example}

\section{Conclusion}\label{s7}

We investigate the opacities of DES under sensor attacks using the ALTER model in this paper. The main contributions of the paper can be summarized as follows. 
(1) We extend the property of opacities to the case with sensor attacks. Specifically, we define opacities for two different settings as $W$-$CA$-opacities and $N$-$CA$-opacities.
(2) By constructing an automaton that integrates the information of the ALTER models, we find an approach to check $W$-$CA$-opacities. 
(3) By the parallel composition on two observers, we obtain a method of verifying $N$-$CA$-opacities. 
(4) By adopting the detector technique, we find polynomial algorithms for verifying weak opacity and non-opacity for both settings. 

In the future, we plan to use $CA$-opacities for information control in DES under sensor attacks.

\bibliographystyle{ieeetr}

\bibliography{CyberAttack}

@INPROCEEDINGS{9501323,
  author={Zheng, Shengbao and Shu, Shaolong and Lin, Feng},
  booktitle={2021 6th International Conference on Automation, Control and Robotics Engineering (CACRE)}, 
  title={Modeling and Control of Discrete Event Systems under Joint Sensor-Actuator Cyber Attacks}, 
  year={2021},
  volume={},
  number={},
  pages={216-220},
  doi={10.1109/CACRE52464.2021.9501323}}

@article{yang2021opacity,
  title={Opacity of networked discrete event systems},
  author={Yang, Jingkai and Deng, Weilin and Qiu, Daowen and Jiang, Cheng},
  journal={Information Sciences},
  volume={543},
  pages={328--344},
  year={2021},
  publisher={Elsevier}
}

@article{chen2022attackable,
  title={Attackable detectability of partially-observed discrete-event systems under sensor attack},
  author={Chen, Qinrui and Su, Rong and Li, Zhiwu},
  journal={IFAC-PapersOnLine},
  volume={55},
  number={28},
  pages={121--127},
  year={2022},
  publisher={Elsevier}
}

@article{zhang2021joint,
  title={Joint state estimation under attack of discrete event systems},
  author={Zhang, Qi and Seatzu, Carla and Li, Zhiwu and Giua, Alessandro},
  journal={IEEE Access},
  volume={9},
  pages={168068--168079},
  year={2021},
  publisher={IEEE}
}

@article{lin2023diagnosability,
  title={Diagnosability of Discrete Event Systems under Sensor Attacks},
  author={Lin, Feng and Lafortune, St{\'e}phane and Wang, Caisheng},
  journal={IFAC-PapersOnLine},
  volume={56},
  number={2},
  pages={3572--3578},
  year={2023},
  publisher={Elsevier}
}

@article{li2022tamper,
  title={Tamper-tolerant diagnosability under bounded or unbounded attacks},
  author={Li, Yuting and Hadjicostis, Christoforos N and Wu, Naiqi},
  journal={IFAC-PapersOnLine},
  volume={55},
  number={28},
  pages={52--57},
  year={2022},
  publisher={Elsevier}
}

@article{alves2022robust,
  title={Robust decentralized diagnosability of networked discrete event systems against DoS and deception attacks},
  author={Alves, Marcos VS and Barcelos, Raphael J and Carvalho, Lilian K and Basilio, Jo{\~a}o C},
  journal={Nonlinear Analysis: Hybrid Systems},
  volume={44},
  pages={101162},
  year={2022},
  publisher={Elsevier}
}

@inproceedings{li2021error,
  title={Error-and tamper-tolerant decentralized diagnosability of discrete event systems under cost constraints},
  author={Li, Yuting and Hadjicostis, Christoforos N and Wu, Naiqi},
  booktitle={2021 European Control Conference (ECC)},
  pages={42--47},
  year={2021},
  organization={IEEE}
}

@article{zhang2023robust,
  title={Robust predictability in discrete event systems under sensor attacks},
  author={Zhang, Qi},
  journal={Frontiers in Physics},
  volume={11},
  pages={1185103},
  year={2023},
  publisher={Frontiers}
}

@article{jacob2016overview,
	title={Overview of discrete event systems opacity: Models, validation, and quantification},
	author={Jacob, Romain and Lesage, Jean-Jacques and Faure, Jean-Marc},
	journal={Annual reviews in control},
	volume={41},
	pages={135--146},
	year={2016},
	publisher={Elsevier}
}

@article{lafortune2018history,
	title={On the history of diagnosability and opacity in discrete event systems},
	author={Lafortune, St{\'e}phane and Lin, Feng and Hadjicostis, Christoforos N},
	journal={Annual Reviews in Control},
	volume={45},
	pages={257--266},
	year={2018},
	publisher={Elsevier}
}

@article{basilio2021analysis,
	title={Analysis and control for resilience of discrete event systems: Fault diagnosis, opacity and cyber security},
	author={Basilio, Jo{\~a}o Carlos and Hadjicostis, Christoforos N and Su, Rong and others},
	journal={Foundations and Trends{\textregistered} in Systems and Control},
	volume={8},
	number={4},
	pages={285--443},
	year={2021},
	publisher={Now Publishers, Inc.}
}

@article{mazare2004using,
	title={Using unification for opacity properties},
	author={Mazar{\'e}, Laurent},
	journal={Proceedings of the 4th IFIP WG1},
	volume={7},
	pages={165--176},
	year={2004}
}

@article{bryans2006opacity,
  title={Opacity generalised to transition systems},
  author={Bryans, Jeremy W and Koutny, Maciej and Mazar{\'e}, Laurent and Ryan, Peter YA},
  journal={International Journal of Information Security},
  volume={7},
  pages={421--435},
  year={2008},
  publisher={Springer}
}

@article{badouel2007concurrent,
	title={Concurrent secrets},
	author={Badouel, Eric and Bednarczyk, Marek and Borzyszkowski, Andrzej and Caillaud, Beno{\^\i}t and Darondeau, Philippe},
	journal={Discrete Event Dynamic Systems},
	volume={17},
	pages={425--446},
	year={2007},
	publisher={Springer}
}

@inproceedings{saboori2007notions,
	title={Notions of security and opacity in discrete event systems},
	author={Saboori, Anooshiravan and Hadjicostis, Christoforos N},
	booktitle={2007 46th IEEE Conference on Decision and Control},
	pages={5056--5061},
	year={2007},
	organization={IEEE}
}

@article{dubreil2010supervisory,
	title={Supervisory control for opacity},
	author={Dubreil, J{\'e}r{\'e}my and Darondeau, Philippe and Marchand, Herv{\'e}},
	journal={IEEE Transactions on Automatic Control},
	volume={55},
	number={5},
	pages={1089--1100},
	year={2010},
	publisher={IEEE}
}

@article{wu2013comparative,
	title={Comparative analysis of related notions of opacity in centralized and coordinated architectures},
	author={Wu, Yi-Chin and Lafortune, St{\'e}phane},
	journal={Discrete Event Dynamic Systems},
	volume={23},
	number={3},
	pages={307--339},
	year={2013},
	publisher={Springer}
}

@article{tong2018current,
	title={Current-state opacity enforcement in discrete event systems under incomparable observations},
	author={Tong, Yin and Li, Zhiwu and Seatzu, Carla and Giua, Alessandro},
	journal={Discrete Event Dynamic Systems},
	volume={28},
	pages={161--182},
	year={2018},
	publisher={Springer}
}

@article{lin2011opacity,
  title={Opacity of discrete event systems and its applications},
  author={Lin, Feng},
  journal={Automatica},
  volume={47},
  number={3},
  pages={496--503},
  year={2011},
  publisher={Elsevier}
}

@article{ben2016opaque,
  title={Opaque superlanguages and sublanguages in discrete event systems},
  author={Ben-Kalefa, Majed and Lin, Feng},
  journal={Cybernetics and Systems},
  volume={47},
  number={5},
  pages={392--426},
  year={2016},
  publisher={Taylor \& Francis}
}

@article{behinaein2019optimal,
	title={Optimal information release for mixed opacity in discrete-event systems},
	author={Behinaein, Behnam and Lin, Feng and Rudie, Karen},
	journal={IEEE Transactions on Automation Science and Engineering},
	volume={16},
	number={4},
	pages={1960--1970},
	year={2019},
	publisher={IEEE}
}

@inproceedings{rashidinejad2019supervisory,
	title={Supervisory control of discrete-event systems under attacks: An overview and outlook},
	author={Rashidinejad, Aida and Wetzels, Bart and Reniers, Michel and Lin, Liyong and Zhu, Yuting and Su, Rong},
	booktitle={2019 18th European Control Conference (ECC)},
	pages={1732--1739},
	year={2019},
	organization={IEEE}
}

@article{wakaiki2019supervisory,
	title={Supervisory control of discrete-event systems under attacks},
	author={Wakaiki, Masashi and Tabuada, Paulo and Hespanha, Jo{\~a}o P},
	journal={Dynamic Games and Applications},
	volume={9},
	pages={965--983},
	year={2019},
	publisher={Springer}
}

@article{alves2022discrete,
	title={Discrete-event systems subject to unknown sensor attacks},
	author={Alves, Michel RC and Pena, Patr{\'\i}cia N and Rudie, Karen},
	journal={Discrete Event Dynamic Systems},
	volume={32},
	number={1},
	pages={143--158},
	year={2022},
	publisher={Springer}
}

@ARTICLE{zheng2021modeling,
  author={Zheng, Shengbao and Shu, Shaolong and Lin, Feng},
  journal={IEEE Transactions on Control of Network Systems}, 
  title={Modeling and Control of Discrete-Event Systems Under Joint Sensor–Actuator Cyberattacks}, 
  year={2024},
  volume={11},
  number={2},
  pages={782-794},
  doi={10.1109/TCNS.2023.3312249}}

@article{shu2011generalized,
	title={Generalized detectability for discrete event systems},
	author={Shu, Shaolong and Lin, Feng},
	journal={Systems \& control letters},
	volume={60},
	number={5},
	pages={310--317},
	year={2011},
	publisher={Elsevier}
}

@book{cassandras2021introduction,
	title={Introduction to Discrete Event Systems},
	author={Cassandras, Christos G. and Lafortune, St\'ephane},
	edition = {3},
	year={2021},
	publisher={Springer Nature}
}

@article{YangJ2021opacity,
  title={Opacity of networked discrete event systems},
  author={Yang, Jingkai and Deng, Weilin and Qiu, Daowen and Jiang, Cheng},
  journal={Information Sciences},
  volume={543},
  pages={328--344},
  year={2021},
  publisher={Elsevier}
}

@inproceedings{yin2018verification,
  title={Verification of opacity in networked supervisory control systems with insecure control channels},
  author={Yin, Xiang and Li, Shaoyuan},
  booktitle={2018 IEEE Conference on Decision and Control (CDC)},
  pages={4851--4856},
  year={2018},
  organization={IEEE}
}

@inproceedings{yin2019opacity,
  title={Opacity of networked supervisory control systems over insecure multiple channel networks},
  author={Yin, Xiang and Li, Shaoyuan},
  booktitle={2019 IEEE 58th Conference on Decision and Control (CDC)},
  pages={7641--7646},
  year={2019},
  organization={IEEE}
}

@article{habbachi2022secret,
  title={Secret inference and attacktability analysis of discrete event systems},
  author={Habbachi, Salwa and Zaghdoud, Amal and Li, Zhiwu and Wu, Naiqi and Khalgui, Mohamed},
  journal={Information Sciences},
  volume={609},
  pages={1221--1238},
  year={2022},
  publisher={Elsevier}
}

@article{meira2020synthesis,
  title={Synthesis of sensor deception attacks at the supervisory layer of cyber--physical systems},
  author={Meira-G{\'o}es, R{\^o}mulo and Kang, Eunsuk and Kwong, Raymond H and Lafortune, St{\'e}phane},
  journal={Automatica},
  volume={121},
  pages={109172},
  year={2020},
  publisher={Elsevier}
}

@article{meira2021synthesis,
  title={Synthesis of Optimal Multiobjective Attack Strategies for Controlled Systems Modeled by Probabilistic Automata},
  author={Meira-G{\'o}es, R{\^o}mulo and Kwong, Raymond H and Lafortune, St{\'e}phane},
  journal={IEEE Transactions on Automatic Control},
  volume={67},
  number={6},
  pages={2873--2888},
  year={2021},
  publisher={IEEE}
}

@article{meira2021synthesisS,
  title={Synthesis of supervisors robust against sensor deception attacks},
  author={Meira-G{\'o}es, R{\^o}mulo and Lafortune, St{\'e}phane and Marchand, Herv{\'e}},
  journal={IEEE Transactions on Automatic Control},
  volume={66},
  number={10},
  pages={4990--4997},
  year={2021},
  publisher={IEEE}
}

@article{lima2021security,
  title={Security of cyber-physical systems: design of a security supervisor to thwart attacks},
  author={Lima, P{\'u}blio M and Alves, Marcos VS and Carvalho, Lilian Kawakami and Moreira, Marcos V},
  journal={IEEE Transactions on Automation Science and Engineering},
  volume={19},
  number={3},
  pages={2030--2041},
  year={2021},
  publisher={IEEE}
}

@article{su2018supervisor,
  title={Supervisor synthesis to thwart cyber attack with bounded sensor reading alterations},
  author={Su, Rong},
  journal={Automatica},
  volume={94},
  pages={35--44},
  year={2018},
  publisher={Elsevier}
}

@article{ji2018enforcement,
  title={Enforcement of opacity by public and private insertion functions},
  author={Ji, Yiding and Wu, Yi-Chin and Lafortune, St{\'e}phane},
  journal={Automatica},
  volume={93},
  pages={369--378},
  year={2018},
  publisher={Elsevier}
}

@article{wu2014synthesis,
  title={Synthesis of insertion functions for enforcement of opacity security properties},
  author={Wu, Yi-Chin and Lafortune, St{\'e}phane},
  journal={Automatica},
  volume={50},
  number={5},
  pages={1336--1348},
  year={2014},
  publisher={Elsevier}
}

@article{wu2015synthesis,
  title={Synthesis of optimal insertion functions for opacity enforcement},
  author={Wu, Yi-Chin and Lafortune, St{\'e}phane},
  journal={IEEE Transactions on Automatic Control},
  volume={61},
  number={3},
  pages={571--584},
  year={2015},
  publisher={IEEE}
}

@article{ji2019opacity,
  title={Opacity enforcement using nondeterministic publicly known edit functions},
  author={Ji, Yiding and Yin, Xiang and Lafortune, St{\'e}phane},
  journal={IEEE Transactions on Automatic Control},
  volume={64},
  number={10},
  pages={4369--4376},
  year={2019},
  publisher={IEEE}
}

@article{mohajerani2019compositional,
  title={Compositional and abstraction-based approach for synthesis of edit functions for opacity enforcement},
  author={Mohajerani, Sahar and Ji, Yiding and Lafortune, St{\'e}phane},
  journal={IEEE Transactions on Automatic Control},
  volume={65},
  number={8},
  pages={3349--3364},
  year={2019},
  publisher={IEEE}
}

@article{zhang2015maximum,
  title={Maximum information release while ensuring opacity in discrete event systems},
  author={Zhang, Bo and Shu, Shaolong and Lin, Feng},
  journal={IEEE Transactions on Automation Science and Engineering},
  volume={12},
  number={3},
  pages={1067--1079},
  year={2015},
  publisher={IEEE}
}

\end{document}